# A Stable and General Quantum Fractional-Step Lattice Boltzmann Method for Incompressible Flows

Yang Xiao[1], Liming Yang[1, 2, 3, *], Chang Shu[4] and Yinjie Du[1]

[1]Department of Aerodynamics, College of Aerospace Engineering, Nanjing University of Aeronautics and Astronautics, Nanjing 210016, China

[2]State Key Laboratory of Mechanics and Control for Aerospace Structures, Nanjing University of Aeronautics and Astronautics, Nanjing 210016, China

[3]MIIT Key Laboratory of Unsteady Aerodynamics and Flow Control, Nanjing University of Aeronautics and Astronautics, Nanjing 210016, China

[4]Department of Mechanical Engineering, National University of Singapore, Singapore 117576, Singapore

## Abstract

Quantum computing shows substantial potential in accelerating simulations and alleviating memory bottlenecks in computational fluid dynamics (CFD), owing to its inherent properties of superposition and entanglement. The lattice Boltzmann method (LBM), being largely algebraic in nature, has inspired the development of various quantum LBMs. However, most existing approaches fix the relaxation time at $\tau = 1$, thereby confining a given mesh resolution to simulations at a single Reynolds number. Although our earlier quantum lattice kinetic scheme (LKS) lifted this restriction, it suffers from instability at high Reynolds numbers. To address this challenge, we propose a quantum fractional-step LBM (FS-LBM). In this framework, the predictor step is implemented on a quantum circuit using the standard LBM formulation, while the corrector step is performed classically. The relaxation time is retained at $\tau = 1$ to ensure seamless compatibility with existing quantum LBMs. Benchmark simulations of representative two- and three-dimensional incompressible isothermal and thermal flows demonstrate that the quantum FS-LBM achieves accuracy and convergence orders consistent with its classical counterpart, while significantly outperforming the quantum LKS in both precision and stability. Notably, this work presents the first quantum LBM simulation of three-dimensional incompressible thermal flows.

---

*Corresponding author, E-mail: lmyang@nuaa.edu.cn.

# 1 Introduction

The lattice Boltzmann method (LBM), derived from kinetic theory, offers advantages such as simple algebraic operations, ease of implementation, and natural parallelism [1-3]. Leveraging these distinctive features, LBM has been successfully applied to various incompressible flows, including multiphase flows [4-6], thermal fluids [7-9], and magnetohydrodynamics [10-12]. However, LBM typically requires storing the distribution functions, resulting in substantially greater memory consumption than traditional Navier-Stokes (N-S) solvers [13-15]. In addition, LBM exhibits poor stability at high Reynolds numbers. As reported in [16-17], the relaxation parameter $\tau = 0.5$ is the margin of instability, while $\tau$ naturally approaches 0.5 if the mesh resolution is insufficient at high Reynolds numbers.

To improve stability and efficiency, several extensions of LBM have been proposed. d'Humiéres et al. [18] introduced the multi-relaxation-time LBM, which enhances the computational stability but increases computational effort. Inamuro [19] proposed the lattice kinetic scheme (LKS), which fixes the relaxation time $\tau$ to 1 and updates macroscopic quantities solely from the equilibrium distribution function (EDF). This reduces memory requirements since only the EDF must be stored. However, Shu et al. [20] pointed out that the LKS suffers from poor numerical stability at high Reynolds number flows. To address this limitation, they proposed the fractional-step LBM (FS-LBM), which significantly enhances numerical stability under high Reynolds number conditions [20]. The FS-LBM consists of a predictor step and a corrector step. In the predictor step, the relaxation time $\tau$ is set to 1 and the standard LBM is applied, allowing the use of only the EDF without modification as in LKS. In the corrector step, an anti-diffusion equation is solved using the central difference scheme. Building on this idea, Chen et al. [21-23] proposed the simplified LBM (SLBM), another fractional-step-based method that enhances stability while still requiring storage of only the EDF. In contrast to FS-LBM, the corrector step in SLBM is carried out using the EDF directly, thereby avoiding the need to solve the anti-diffusion equation. Nevertheless, even storing only the EDF still results in significant memory requirements, particularly for large-scale engineering applications.

Recently, quantum computing has emerged as a potential solution to the significant memory demands of LBM due to its unique properties of superposition and entanglement [24-25]. A quantum bit (qubit) can exist in a superposition of 0 and 1 states, enabling an $n$-qubit system to represent a linear combination of $2^n$ basis states [26-27]. For instance, storing flow variables over $10^{18}$ mesh points using 64-bit precision would require only about 60 ($\sim \log_2 10^{18}$) qubits, whereas a classical computer would need nearly 8000 petabytes, which is far beyond the capacity of most existing classical systems [28-29]. Because LBM involves primarily algebraic operations rather than discrete differential operators, it is particularly suitable for quantum implementation. Several quantum LBMs have been developed. Budinski [30] proposed a quantum LBM for solving 1D and 2D linear advection-diffusion equations by implementing the collision and streaming steps via the linear combination of unitaries (LCU) decomposition [31] and quantum walks [32]. Wawrzyniak et al. [33] extended quantum LBM to 3D advection-diffusion problems using a novel data encoding strategy [34] that significantly reduces the number of quantum gates required for initialization. Budinski [35] further adopted quantum LBM to simulate incompressible flows by employing the vorticity-stream function formulation of the N-S equations. Kocherla et al. [36] proposed a two-circuit quantum LBM, in which two separate quantum circuits are used to compute the vorticity and stream functions, respectively, thereby lowering the quantum gate count relative to single-circuit implementations.

Despite these advances, quantum LBMs face a fundamental limitation, i.e., the computational process in quantum systems is inherently linear, whereas the collision term in LBM is nonlinear [37]. To circumvent this difficulty, most existing quantum LBMs [30, 33, 35-36] set $\tau = 1$, effectively removing nonlinearity. However, this restriction ties each mesh resolution to a single Reynolds number, limiting the flexibility of quantum LBMs. Although the Carleman linearization method can, in principle, eliminate the nonlinearity of the collision term, it leads to exponential growth in system dimensionality [38-39]. In our previous work, we proposed a quantum LKS based on Inamuro's method to simulate 2D and 3D incompressible flows [40]. This scheme retains $\tau = 1$ while allowing the simulation of flows at arbitrary Reynolds numbers, achieving accuracy comparable to its classical counterpart. However, as mentioned earlier [20], LKS suffers from poor stability at high Reynolds numbers, and this limitation persists in the quantum LKS. This highlights the urgent need for a more practical and robust quantum LBM.

In this work, to balance simplicity and broad applicability, we develop a quantum FS-LBM based on the framework of Shu et al. [20] for simulating incompressible flows at arbitrary Reynolds numbers. Specifically, the quantum FS-LBM consists of a predictor step and a corrector step. In the predictor step, the relaxation time $\tau$ is set to 1, and the evolution is implemented via quantum circuits. The corrector step is performed classically by solving the reverse diffusion equation using finite difference methods. This design allows seamless integration with existing approaches [30, 33, 35-36]. To address instabilities that may arise in regions with steep gradients during the anti-diffusion correction, we employ a finite difference scheme with a stable stencil based on least-squares quadratic fitting [41-42], ensuring robustness. Furthermore, we consider two variants of the quantum FS-LBM. The first, referred to as quantum FS-LBM-I, implements collision, streaming, and macroscopic variable computation entirely within quantum circuits. However, due to the limitation that quantum circuits can directly extract only zeroth-order moments, quantum FS-LBM-I requires multiple identical quantum circuits to compute macroscopic density and velocity separately, which leads to considerable computational overhead. The second variant, termed quantum FS-LBM-II, executes only collision and streaming steps on the quantum circuit and outputs the post-streaming distribution functions, with macroscopic quantities then computed classically. Since quantum FS-LBM-II requires only a single quantum circuit to obtain both density and velocity, it offers a substantial improvement in computational efficiency. The proposed method is validated through simulations of 2D and 3D Taylor-Green vortex flows, lid-driven cavity flows, and natural convection in a square cavity, confirming its accuracy and effectiveness. Comparative results further show that the quantum FS-LBM exhibits superior accuracy and stability over the quantum LKS. Notably, this work represents the first application of quantum LBM to the simulation of 3D incompressible thermal flows.

The structure of this paper is arranged as follows: Section 1 is the introduction; Section 2 and Section 3 detail the fundamentals of classical FS-LBM and quantum FS-LBM, respectively; Section 4 applies the developed method to solve a series of 2D and 3D numerical cases; Section 5 gives a brief summarization.

## 2 Classical fractional-step lattice Boltzmann method

For incompressible thermal flows, the macroscopic governing equations are typically composed of the weakly compressible N-S equations coupled with the energy equation, which can be expressed as

$$\frac{\partial \rho}{\partial t} + \nabla \cdot (\rho \boldsymbol{u}) = 0,$$
$$\frac{\partial \rho \boldsymbol{u}}{\partial t} + \nabla \cdot (\rho \boldsymbol{u}\boldsymbol{u}) = -\nabla p + \upsilon \nabla^2 \boldsymbol{u} + \boldsymbol{F}, \qquad (1)$$
$$\frac{\partial T}{\partial t} + \nabla \cdot (T\boldsymbol{u}) = \kappa \nabla^2 T,$$

where $\boldsymbol{u}$, $p$, $\rho$, and $T$ denote the velocity, pressure, density, and temperature of fluid flow, respectively. $\upsilon$ and $\kappa$ represent the viscosity and thermal diffusivity. The term $\boldsymbol{F}$ is the buoyancy force caused by temperature non-uniformity, which can be expressed under the Boussinesq approximation as:

$$\boldsymbol{F} = -\rho g \beta (T - T_m) \boldsymbol{j}, \qquad (2)$$

where $g$, $\beta$, $T_m$, and $\boldsymbol{j}$ denote the gravitational acceleration, thermal expansion coefficient, mean temperature of the flow field, and the unit vector in the $y$-direction, respectively. As shown in Eq. (1), the temperature field $T$ is influenced by the flow field, while the buoyancy force $\boldsymbol{F}$ generated by the temperature field $T$ also plays a crucial role as an external force in the momentum equation.

To solve the incompressible thermal flow problem described by Eq. (1) using the LBM, the density distribution function $f$ and the temperature distribution function $h$ are introduced to represent the flow field and temperature field, respectively. The corresponding governing equations are given as follows

$$f_\alpha(\mathbf{x}+\mathbf{e}_\alpha\delta t,t+\delta t)=\left(1-\frac{1}{\tau_\upsilon}\right)f_\alpha(\mathbf{x},t)+\frac{1}{\tau_\upsilon}f_\alpha^{eq}(\mathbf{x},t),$$

$$h_\alpha(\mathbf{x}+\mathbf{e}_\alpha\delta t,t+\delta t)=\left(1-\frac{1}{\tau_\kappa}\right)h_\alpha(\mathbf{x},t)+\frac{1}{\tau_\kappa}h_\alpha^{eq}(\mathbf{x},t),$$

(3)

where $\delta t$ and $\mathbf{e}_\alpha$ are the time step and the discrete particle velocity vector in the $\alpha$ direction. $\tau_\upsilon$ and $\tau_\kappa$ denote the relaxation time of the fluid field and temperature field, respectively. The corresponding EDFs $f_\alpha^{eq}$ and $h_\alpha^{eq}$ are expressed as

$$f_\alpha^{eq}(\mathbf{x},t)=w_\alpha\rho\left(1+\frac{\mathbf{e}_\alpha\cdot\mathbf{u}}{c_s^2}+\frac{(\mathbf{e}_\alpha\cdot\mathbf{u})^2}{2c_s^4}-\frac{\mathbf{u}\cdot\mathbf{u}}{2c_s^2}\right),$$

$$h_\alpha^{eq}(\mathbf{x},t)=w_\alpha T\left(1+\frac{\mathbf{e}_\alpha\cdot\mathbf{u}}{c_s^2}+\frac{(\mathbf{e}_\alpha\cdot\mathbf{u})^2}{2c_s^4}-\frac{\mathbf{u}\cdot\mathbf{u}}{2c_s^2}\right),$$

(4)

where $\rho$, $c_s$, and $w_\alpha$ are the density, lattice sound speed, and weighting coefficient, respectively. The macroscopic quantities, including density, velocity, and temperature, can then be evaluated from the distribution functions according to

$$\rho=\sum_\alpha f_\alpha^{eq}(\mathbf{x},t),\quad \rho\mathbf{u}=\sum_\alpha \mathbf{e}_\alpha f_\alpha^{eq}(\mathbf{x},t),\quad T=\sum_\alpha h_\alpha^{eq}(\mathbf{x},t). \qquad (5)$$

In FS-LBM developed by Shu et al. [20], both relaxation parameters $\tau_\upsilon$ and $\tau_\kappa$ are fixed at 1, thereby simplifying the collision and streaming steps to

**Collision step:**
$$f_\alpha^*(\mathbf{x},t)=f_\alpha^{eq}(\mathbf{x},t),$$
$$h_\alpha^*(\mathbf{x},t)=h_\alpha^{eq}(\mathbf{x},t).$$

**Streaming step:**
$$f_\alpha(\mathbf{x},t+\delta t)=f_\alpha^*(\mathbf{x}-\mathbf{e}_\alpha\delta_t,t),$$
$$h_\alpha(\mathbf{x},t+\delta t)=h_\alpha^*(\mathbf{x}-\mathbf{e}_\alpha\delta_t,t).$$

(6)

where $f_\alpha^*$ and $h_\alpha^*$ are the post-collision states. As observed from Eq. (6), the update of the entire lattice Boltzmann equation (LBE) depends solely on the EDFs, significantly reducing memory requirements since only the EDFs need to be stored. However, according to the Chapman–Enskog expansion analysis, the macroscopic equations recovered from Eq. (6) are as follows

$$\begin{aligned}
&\frac{\partial \rho}{\partial t}+\nabla \cdot(\rho \boldsymbol{u})=0, \\
&\frac{\partial \rho \boldsymbol{u}}{\partial t}+\nabla \cdot(\rho \boldsymbol{u}\boldsymbol{u})=-\nabla p+\upsilon^{*}\nabla^{2}\boldsymbol{u}, \\
&\frac{\partial T}{\partial t}+\nabla \cdot(T\boldsymbol{u})=\kappa^{*}\nabla^{2}T,
\end{aligned} \quad (7)$$

where $\upsilon^{*}=0.5c_{s}^{2}\delta t=\Delta x^{2}/6\delta t$, $\kappa^{*}=0.5c_{s}^{2}\delta t=\Delta x^{2}/6\delta t$. In standard LBM, the lattice spacing $\Delta x$ and time step $\delta t$ are typically set to 1, leading to fixed dynamic viscosity $\upsilon^{*}$ = 1/6 and thermal diffusivity $\kappa^{*}$ = 1/6, which means that a given mesh resolution can only simulate flows at a specific Reynolds number.

To overcome the above limitation, Shu et al. [20] introduced a fractional-step scheme, decomposing FS-LBM into the following predictor and corrector steps

**Predictor step:**       **Corrector step:**

$$\begin{aligned}
&\frac{\partial \rho}{\partial t}=-\nabla \cdot(\rho \boldsymbol{u})=L_{1}^{c}, &&\frac{\partial \rho}{\partial t}=0=L_{2}^{c}, \\
&\frac{\partial \rho \boldsymbol{u}}{\partial t}=-\nabla \cdot(\rho \boldsymbol{u}\boldsymbol{u})-\nabla p+\upsilon^{*}\nabla^{2}\boldsymbol{u}=L_{1}^{m}, &&\frac{\partial \rho \boldsymbol{u}}{\partial t}=(\upsilon-\upsilon^{*})\nabla^{2}\boldsymbol{u}+\boldsymbol{F}=L_{2}^{m}, \\
&\frac{\partial T}{\partial t}=-\nabla \cdot(T\boldsymbol{u})+\kappa^{*}\nabla^{2}T=L_{1}^{T}. &&\frac{\partial T}{\partial t}=(\kappa-\kappa^{*})\nabla^{2}T=L_{2}^{T}.
\end{aligned} \quad (8)$$

In the predictor step, the macroscopic variables $\rho^{n}$, $\boldsymbol{u}^{n}$, and $T^{n}$ at time step $n$ are used to obtain intermediate macroscopic variables $\bar{\rho}$, $\bar{\boldsymbol{u}}$, and $\bar{T}$ according to Eq. (6). In the subsequent corrector step, the time derivatives are discretized via a first-order forward Euler scheme, yielding the updated macroscopic variables $\rho^{n+1}$, $\boldsymbol{u}^{n+1}$, and $T^{n+1}$

$$\begin{aligned}
&\rho^{n+1}=\bar{\rho}, \\
&\rho^{n+1}\boldsymbol{u}^{n+1}=\overline{\rho\boldsymbol{u}}+\delta t(\upsilon-\upsilon^{*})\nabla^{2}\boldsymbol{u}^{n}+\delta t\boldsymbol{F}^{n}, \\
&T^{n+1}=\bar{T}+\delta t(\kappa-\kappa^{*})\nabla^{2}T^{n}.
\end{aligned} \quad (9)$$

For the Laplacian operator $\nabla^{2}$ in Eq. (9), Shu et al. [20] employed the central difference (CD) stencil to discretize it. However, since $\upsilon - \upsilon^{*}$ and $\kappa - \kappa^{*}$ are generally negative, this approach can become unstable in flows with large gradients. To address this, Dupuy et al. [41] proposed a finite difference stable stencil (SS) based on least-squares quadratic fitting, with the discrete 2D Laplacian operator $\nabla^2$ given by

$$\nabla^2 \boldsymbol{u}^n = \frac{2\left(\boldsymbol{u}_{i+1,j+1}^n + \boldsymbol{u}_{i+1,j-1}^n + \boldsymbol{u}_{i-1,j+1}^n + \boldsymbol{u}_{i-1,j-1}^n\right) - \left(\boldsymbol{u}_{i,j+1}^n + \boldsymbol{u}_{i,j-1}^n + \boldsymbol{u}_{i-1,j}^n + \boldsymbol{u}_{i+1,j}^n + 4\boldsymbol{u}_{i,j}^n\right)}{3h^2}$$

$$\nabla^2 T^n = \frac{2\left(T_{i+1,j+1}^n + T_{i+1,j-1}^n + T_{i-1,j+1}^n + T_{i-1,j-1}^n\right) - \left(T_{i,j+1}^n + T_{i,j-1}^n + T_{i-1,j}^n + T_{i+1,j}^n + 4T_{i,j}^n\right)}{3h^2} \quad (10)$$

where $i$ and $j$ correspond to the indices for the $x$ and $y$ directions, respectively. For the 3D Laplacian operator, its discrete form is obtained by applying Eq. (10) independently to each coordinate plane. The SS stencil retains second-order accuracy like the CD stencil, but with reduced accuracy at the same mesh size. Thus, to preserve solution fidelity, the SS stencil is employed only when the CD stencil becomes unstable.

Notably, when using the FS-LBM to simulate incompressible isothermal flows, Eq. (6) involves only the density distribution function $f_\alpha$, and the buoyancy term $\boldsymbol{F}$ in Eq. (8) vanishes.

## 3 Quantum fractional-step lattice Boltzmann method

As outlined in Section 2, the classical FS-LBM consists of a predictor step and a corrector step. In the quantum FS-LBM, the predictor step is implemented through a quantum circuit, whereas the corrector step is performed on a classical computer. Furthermore, the evolution of the density and temperature distribution functions is realized through distinct but structurally identical quantum circuits. For clarity, we illustrate the quantum FS-LBM using only the density distribution function.

As illustrated in Fig. 1(a), the circuit of quantum FS-LBM is composed of five quantum registers and four core modules: initialization, collision, streaming, and computation of macroscopic variables. The qubit registers $|q_d\rangle$ with $d = \{x, y, z\}$ encode the variables at the lattice points, each requiring $n_d = \log_2(\{M_x, M_y, M_z\})$ qubits, where $M_x$, $M_y$, $M_z$ are the number of lattice points in each spatial direction. In the LBM framework, the distribution function for each discrete velocity direction involves the variables at the same lattice point, while the qubit register $|q_d\rangle$ can only store the variables for one discrete velocity direction. Therefore, the qubit register $|q_Q\rangle$ is introduced to store the variables at the lattice points corresponding to each discrete velocity direction $\alpha$, which requires $n_Q = \text{ceil}(\log_2 Q)$ qubits, where $Q$ is the number of discrete velocity directions in the lattice Boltzmann model. An auxiliary register $|a_0\rangle$, consisting of a single qubit ($n_a = 1$), is used for the implementation of the collision and macroscopic variable calculation modules. Consequently, the total number of qubits required for the quantum FS-LBM circuit is given by $n_{total} = \sum_{i \in d} n_i + n_Q + n_a$. In this work, we employ the D2Q9 and D3Q27 lattice models to simulate 2D and 3D incompressible flows, respectively. It is worth noting that for 2D simulations, the qubit register $|q_z\rangle$ can be omitted without loss of generality.

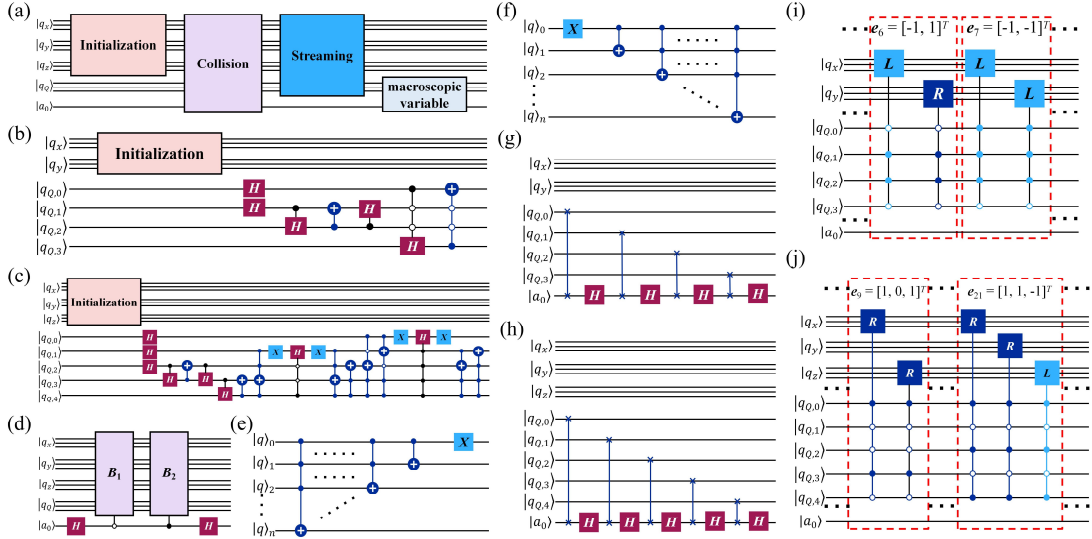

**Fig. 1 Schematic diagram of quantum FS-LBM.** (a) Quantum circuit with general blocks for quantum FS-LBM. Initialization circuits for (b) D2Q9 and (c) D3Q27 lattice models. (d) Quantum circuits for the collision operator. (e) Quantum circuits for the streaming operator $P$. (f) Quantum circuits for the streaming operator $N$. Quantum circuits for the macroscopic variables computation using (g) D2Q9 and (h) D3Q27 lattice model. (i) Quantum circuits for the streaming step in selected directions $e_6 = [-1, 1]^T$ and $e_7 = [-1, -1]^T$ of the D2Q9 lattice model. (j) Quantum circuits for the streaming step in selected directions $e_9 = [-1, 0, 1]^T$ and $e_{21} = [1, 1, -1]^T$ of the D3Q27 lattice model.

**3.1 Initialization step with duplication sequence**

In the initialization step, the macroscopic density field is encoded into the amplitudes of a quantum state vector using the amplitude encoding method proposed by Shende et al. [34], which has already been integrated in the Qiskit platform [43]. In existing quantum LBMs [30, 35-36], all discrete velocity directions are encoded simultaneously, resulting in a computational complexity of $O\left(Q \cdot 2^{\sum_{i \in d} n_i}\right)$. However, since the density field $\rho$ is identical across all discrete velocity directions, we adopt a duplication-based initialization strategy [33, 40] that reduces the computational cost to $O\left(2^{\sum_{i \in d} n_i}\right)$, as shown in Figs 1(b) and 1(c). Specifically, the amplitude encoding is first applied to the density field of a single discrete velocity direction, yielding the quantum state $|\psi_A\rangle = |0\rangle_a |0000\rangle_Q \frac{1}{\|\boldsymbol{\rho}\|} \sum_{k=0}^{2^{N_{init}}-1} \rho_k |k\rangle$, where $N_{init} = \sum_{i \in d} n_i$, and $\|\bullet\|$ is the Euclidean norm. Using a sequence of Hadamard and controlled gates, this encoded density field $\rho$ is then duplicated across all discrete velocity directions.

Taking the D2Q9 lattice model as an example (Fig. 1(b)), the Hadamard gates are applied to the first two qubits of the $|q_Q\rangle$ register, resulting in the evolution of the state to:

$$|\psi_B\rangle = |0\rangle_a \left(\frac{1}{2}|0000\rangle_Q + \frac{1}{2}|0001\rangle_Q + \frac{1}{2}|0010\rangle_Q + \frac{1}{2}|0011\rangle_Q\right) \frac{1}{\|\boldsymbol{\rho}\|} \sum_{k=0}^{2^{N_{init}}-1} \rho_k |k\rangle. \quad (11)$$

By applying subsequent controlled gates, the quantum state further evolves to:

$$\begin{aligned}|\psi_C\rangle = |0\rangle_a (&\frac{1}{2}|0000\rangle_Q + \frac{1}{2\sqrt{2}}|0001\rangle_Q + \frac{1}{2\sqrt{2}}|0010\rangle_Q + \frac{1}{2\sqrt{2}}|0011\rangle_Q \\ &+ \frac{1}{4}|0100\rangle_Q + \frac{1}{4}|0101\rangle_Q + \frac{1}{4}|0110\rangle_Q + \frac{1}{4}|0111\rangle_Q + \frac{1}{2\sqrt{2}}|1000\rangle_Q) \frac{1}{\|\boldsymbol{\rho}\|} \sum_{k=0}^{2^{N_{init}}-1} \rho_k |k\rangle.\end{aligned} \quad (12)$$

Clearly, each subspace of the qubit register $|q_Q\rangle$ represents the density field corresponding to a distinct discrete velocity direction. For the D3Q27 lattice model, the coefficients of each subspace can be derived similarly, with details provided in [33, 40].

## 3.2 Collision step based on the LCU method

As discussed in Section 2, when the relaxation time is set to 1, the collision step in FS-LBM essentially reduces to computing the EDF (Eq. (6)). Since the initialization step encodes only the density field $\rho$, the collision step is implemented via the following collision matrix $D$,

$$D = \begin{bmatrix} C_0 D_0 & 0 & 0 \\ 0 & \ddots & 0 \\ 0 & 0 & C_{Q-1} D_{Q-1} \end{bmatrix}, \quad D_\alpha = \begin{bmatrix} f_\alpha^{eq}(x_0,t) & 0 & 0 \\ 0 & \ddots & 0 \\ 0 & 0 & f_\alpha^{eq}(x_{M-1},t) \end{bmatrix}, \quad (13)$$

where

$$f_\alpha^{eq}(x,t) = f_\alpha^{eq}(x,t)/\rho = w_\alpha \left(1 + \frac{e_\alpha \cdot u}{c_s^2} + \frac{(e_\alpha \cdot u)^2}{2c_s^4} - \frac{u \cdot u}{2c_s^2}\right). \quad (14)$$

Here, $C_\alpha = \sqrt{2}^h$ is a constant determined by the replication sequence in the initialization step, and $h$ represents the number of Hadamard gates applied to the scalar field. Obviously, applying the collision matrix $D$ to the quantum state $|\psi_C\rangle$ yields

$$D \cdot |\psi_C\rangle = |0\rangle_a \sum_{\alpha=0}^{Q} \sum_{k=1}^{2^{N_{init}}-1} \frac{1}{\|\rho\|} f_\alpha^{eq}(x_k,t) |\alpha\rangle_Q |k\rangle \quad (15)$$

which contains the full EDFs and corresponds to the post-collision state $f_\alpha^*(x,t)$ in Eq. (6). However, although the collision matrix $D$ is diagonal, it is not unitary. Hence, it must be decomposed into a linear combination of unitary operators via the linear combination of unitaries (LCU) method:

$$D = (B_1 + B_2)/2, \quad B_{1,2} = D \pm i\sqrt{I - D^2}. \quad (16)$$

As shown in Fig. 1(d), the complete collision step can be realized by applying the Hadamard gates together with two controlled diagonal unitary operators, $B_1$ and $B_2$.

The resulting collision operator can be written as

$$\left(H^* \otimes I^{\otimes n_{total}-1}\right)\left(|0\rangle\langle 0|_a \otimes B_1 + |1\rangle\langle 1|_a \otimes B_2\right)\left(H \otimes I^{\otimes n_{total}-1}\right). \tag{17}$$

Applying the operator in Eq. (17) to the quantum state $|\psi_C\rangle$ in Eq. (12) yields

$$\begin{aligned}|\psi_D\rangle &= |0\rangle_a D|\psi_C\rangle + |1\rangle_a \left(i\sqrt{I-D^2}\right)|\psi_C\rangle \\ &= |0\rangle_a \frac{1}{\|\boldsymbol{\rho}\|}\sum_{\alpha=0}^{Q-1}\sum_{k=0}^{2^{N_{init}}-1} f_\alpha^{eq}(\boldsymbol{x}_k,t)|\alpha\rangle_Q|k\rangle + |1\rangle_a \left(i\sqrt{I-D^2}\right)|\psi_C\rangle.\end{aligned} \tag{18}$$

From Eq. (18), the $|0\rangle_a$ state of the qubit register $|a_0\rangle$ contains the distribution functions $f_\alpha^{eq}(\boldsymbol{x},t)$ for all discrete velocity directions, thus completing the collision step (Eq. (6)).

### 3.3 Streaming step based on the quantum walk

The streaming step can be realized via quantum walks [30, 32]. As shown in Eq. (6), this step essentially corresponds to propagating the post-collision distribution function $f_\alpha^*(\boldsymbol{x},t)$ along its discrete velocity direction $\boldsymbol{e}_\alpha$ to the neighboring lattice sites. This process can be constructed by introducing the positive shift operator $\boldsymbol{P}$ and the negative shift operator $\boldsymbol{N}$, whose matrix forms are given by

$$\boldsymbol{P} = \begin{bmatrix} 0 & 0 & 0 & \cdots & 0 & 1 \\ 1 & 0 & 0 & \cdots & 0 & 0 \\ 0 & 1 & 0 & \cdots & 0 & 0 \\ \vdots & \vdots & \vdots & \ddots & \vdots & \vdots \\ 0 & 0 & 0 & \cdots & 0 & 0 \\ 0 & 0 & 0 & \cdots & 1 & 0 \end{bmatrix}, \quad \boldsymbol{N} = \begin{bmatrix} 0 & 1 & 0 & \cdots & 0 & 0 \\ 0 & 0 & 1 & \cdots & 0 & 0 \\ 0 & 0 & 0 & \cdots & 0 & 0 \\ \vdots & \vdots & \vdots & \ddots & \vdots & \vdots \\ 0 & 0 & 0 & \cdots & 0 & 1 \\ 1 & 0 & 0 & \cdots & 0 & 0 \end{bmatrix}. \tag{19}$$

The quantum circuits for operators $\boldsymbol{P}$ and $\boldsymbol{N}$ are shown in Fig. 1(e) and 1(f), which are composed of multi-controlled X gates.

The shift operator acts solely on the $|q_d\rangle$ register, while the $|q_Q\rangle$ register provides control conditions to guarantee that the streaming step is executed exclusively within the subspace associated with the selected discrete velocity direction. Figs. 1(i) and 1(j) illustrate the quantum circuits of the streaming step for the D2Q9 and D3Q27 lattice models, with representative controlled-shift circuits detailed for specific discrete velocity directions. When the quantum circuits in Figs. 1(i) and 1(j) are applied to the quantum state $|\psi_D\rangle$, the system evolves into the following state

$$|\psi_E\rangle = |0\rangle_a \frac{1}{\|\boldsymbol{\rho}\|} \sum_{\alpha=0}^{Q-1} \sum_{k=0}^{2^{N_{init}}-1} f_\alpha^*(\boldsymbol{x}_k - \boldsymbol{e}_\alpha \delta t, t) |\alpha\rangle_Q |k\rangle + |1\rangle_a |\psi^*\rangle, \quad (20)$$

where $|1\rangle_a |\psi^*\rangle$ denotes the orthogonal state of no further interest. As shown in Eq. (20), the post-streaming distribution function $f_\alpha^*(\boldsymbol{x}_k - \boldsymbol{e}_\alpha \delta t, t)$ is contained in the $|0\rangle_a$ state of the $|a_0\rangle$ register. Importantly, the shift operator inherently satisfies periodic boundary conditions, thereby eliminating the need for additional boundary treatment in flows with periodic boundaries.

### 3.4 Computation of macroscopic variables

The macroscopic density is obtained by computing the 0th moment of the distribution function, which involves summing the distribution function over all discrete velocity directions, as expressed in Eq. (5). In the quantum circuit, this process can be realized by sequentially applying swap gates and Hadamard gates on the $|q_Q\rangle$ register and the auxiliary $|a_0\rangle$ qubit. The swap gates reorder the distribution function, while the Hadamard gates generate amplitude superposition. Fig. 1(g) and 1(h) illustrate the corresponding quantum circuits for computing macroscopic density using the D2Q9 and D3Q27 lattice models, respectively. It can be observed that the circuit complexity depends only on the number of qubits in the $|q_Q\rangle$ register. In other words, it is solely determined by the number of discrete velocity directions and is independent of the mesh size. Applying the quantum circuits in Figs. 1(g) and 1(h) to the quantum state $|\psi_E\rangle$, followed by measuring the auxiliary $|a_0\rangle$ register in the $|0\rangle_a$ state and taking the real part yields

$$\boldsymbol{\rho}^* = 1\Big/\left(\sqrt{2}^{n_Q}\|\boldsymbol{\rho}\|\right)\sum_{\alpha=0}^{Q-1} f_\alpha^*(\boldsymbol{x}-\boldsymbol{e}_\alpha\delta t, t) = 1\Big/\left(\sqrt{2}^{n_Q}\|\boldsymbol{\rho}\|\right)\boldsymbol{\rho}(\boldsymbol{x}, t+\delta t). \tag{21}$$

The density field $\boldsymbol{\rho}^*$ derived from Eq. (21) is then scaled by a constant $\sqrt{2}^{n_Q}\|\boldsymbol{\rho}\|$, yielding the density field $\boldsymbol{\rho}(\boldsymbol{x}, t+\delta t)$ at the next time step $t+\delta t$.

For the calculation of the temperature field *T*, a quantum circuit of the same structure is employed, using the temperature field as input. However, calculating the velocity field $\boldsymbol{u}$ requires the 1st moment of the distribution function $f_\alpha$, whereas the circuits shown in Fig. 1(g) and 1(h) are limited to computing only the 0th moment. To obtain the velocity field, we construct multiple quantum circuits of identical structure, with the product of particle velocity and density field $e_\alpha\boldsymbol{\rho}$ serving as the input, as illustrated in Fig. 2(a). This design allows both quantities to be extracted in parallel from the 0th moment, and we refer to this scheme as quantum FS-LBM-I. However, for 3D incompressible thermal flows, the quantum FS-LBM-I requires five identical quantum circuits, leading to a considerable demand for quantum resources. To address this, we propose an alternative scheme, termed quantum FS-LBM-II, as shown in Fig. 2(b). Unlike quantum FS-LBM-I, quantum FS-LBM-II utilizes only two circuits, one for the flow field and one for the temperature field. In the flow-field circuit, the computation of the macroscopic quantities is omitted, yielding only the post-streaming distribution function $f_\alpha^*(x-e_\alpha\delta t,t)$, while the macroscopic quantities are subsequently evaluated on a classical computer. This hybrid design significantly reduces quantum resource consumption while improving computational efficiency.

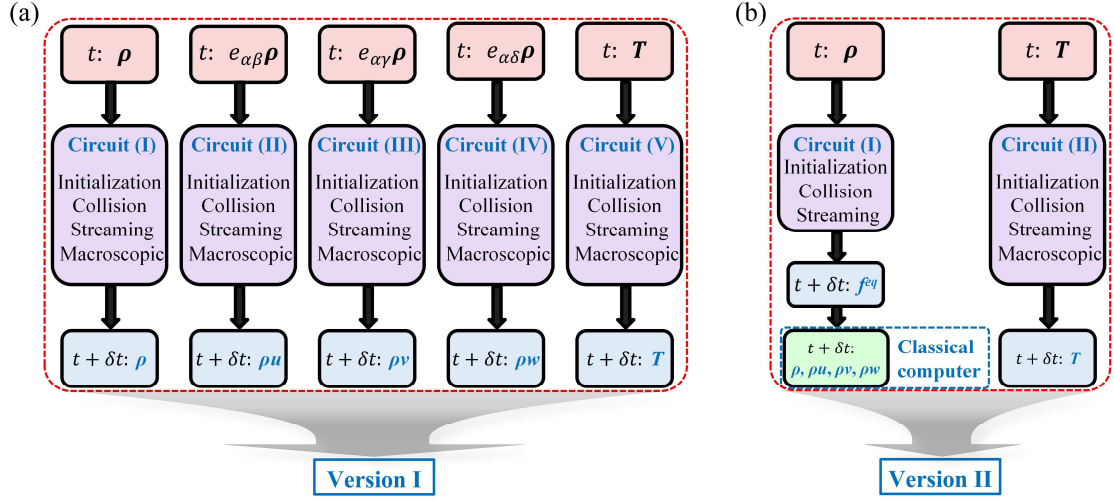

**Fig. 2** Schematic of (a) quantum FS-LBM-I and (b) quantum FS-LBM-II for simulating incompressible flows.

## 4 Numerical Examples

In this section, we apply the proposed quantum FS-LBMs to solve 2D and 3D incompressible flows at various Reynolds and Rayleigh numbers. All simulations are carried out on a PC with an Intel i7-14700KF CPU. The implementation is based on the Qiskit SDK version 0.45.2 and the AER backend version 0.13.2 [43]. For validation, the results of the quantum FS-LBMs are compared with those obtained from the quantum lattice kinetic scheme (LKS) [40]. Details of the quantum LKS for incompressible thermal flows are provided in Appendix A. For steady-state flows, computational convergence is assessed using the residual criterion

$$\sqrt{\frac{\left(\boldsymbol{u}^{n+1}-\boldsymbol{u}^{n}\right)^{2}+\left(T^{n+1}-T^{n}\right)^{2}}{\left(\boldsymbol{u}^{n+1}\right)^{2}+\left(T^{n+1}\right)^{2}}}<\varepsilon, \tag{21}$$

where $\boldsymbol{u}^{n+1}$, $T^{n+1}$, and $\boldsymbol{u}^{n}$, $T^{n}$ denote the velocity and temperature at the time steps $n+1$ and $n$, respectively. Convergence is assumed once the residual falls below $\varepsilon = 10^{-6}$ or $10^{-7}$, at which point the simulation is terminated.

**4.1 Two- and three-dimensional Taylor-Green vortex flows**

We first evaluate the performance and convergence order of the proposed quantum FS-LBMs using 2D and 3D Taylor-Green vortex flows, whose analytical solutions are given by

$$\begin{cases} u(\boldsymbol{x},t)=-u_{0}\cos\left(\frac{\pi x}{L}\right)\sin\left(\frac{\pi y}{L}\right)e^{\frac{-2\pi^{2}u_{0}t}{ReL}}, \\ v(\boldsymbol{x},t)=u_{0}\sin\left(\frac{\pi x}{L}\right)\cos\left(\frac{\pi y}{L}\right)e^{\frac{-2\pi^{2}u_{0}t}{ReL}}, \\ w(\boldsymbol{x},t)=0, \\ \rho(\boldsymbol{x},t)=\rho_{0}-\frac{\rho_{0}u_{0}^{2}}{4c_{s}^{2}}\left[\cos\left(\frac{2\pi x}{L}\right)+\cos\left(\frac{2\pi y}{L}\right)\right]e^{\frac{-4\pi^{2}u_{0}t}{ReL}}. \end{cases} \tag{22}$$

where $u_0$, $\rho_0$, and $Re$ are the characteristic velocity, reference density, and Reynolds number, respectively. The computational domains are specified as $[-L, L] \times [-L, L]$ for 2D and $[-L, L] \times [-L, L] \times [-L, L]$ for 3D cases. Four different uniform meshes ($N \times N$ or $N \times N \times N$, $N = 8, 16, 32,$ and $64$) are used for simulation. The Reynolds number is defined as $Re = u_0 L/\upsilon$. Periodic boundary conditions are applied in all boundaries, and the initial conditions are provided by the analytical solution. Simulation parameters are fixed as $\delta_t = 1$, $\Delta x = 1$, $\rho_0 = 1$, and $Re = 10$. At the dimensionless time $t^* = u_0 t/L = 1.0$, the numerical solution is compared with the analytical one. The relative error is quantified using the $L_2$-norm, defined as $L_2 = \sqrt{\sum[(u_{\text{numerical}} - u_{\text{exact}})/u_0]^2 / N_{\text{total}}}$, where $u_{\text{numerical}}$, $u_{\text{exact}}$, and $N_{\text{total}}$ are the numerical solution, analytical solution, and the total number of mesh points, respectively.

Figs. 3(a) and 3(b) respectively show the relationship between the $L_2$-norm and mesh size on a log scale for 2D and 3D Taylor-Green vortex flows, comparing classical and quantum FS-LBMs with classical and quantum LKSs. It can be seen that both classical and quantum FS-LBMs exhibit convergence rates close to second order in the 2D and 3D Taylor-Green vortex flows, consistent with the theoretical accuracy of LBM. For the same mesh size, the $L_2$-norms of the classical FS-LBM, quantum FS-LBM-I, and quantum FS-LBM-II coincide, and are 1-2 orders of magnitude smaller than those of the classical and quantum LKSs, confirming that FS-LBM provides more accurate results than LKS. Furthermore, Figs. 4 and 5 illustrate the $u$- and $v$-velocity contours of the 2D and 3D Taylor-Green vortices at a mesh size of $N = 64$, obtained using the classical FS-LBM, quantum FS-LBM-I, and quantum FS-LBM-II. The results of the three approaches are in perfect agreement, thereby validating both the accuracy and the effectiveness of the quantum FS-LBMs.

Table 1 summarizes the computational times of quantum FS-LBM-I, quantum FS-LBM-II, quantum LKS-I, and quantum LKS-II at different mesh sizes. For the same version and mesh size, the computational times of quantum FS-LBMs and quantum LKSs are nearly identical, that is, quantum FS-LBM-I versus quantum LKS-I, and quantum FS-LBM-II versus quantum LKS-II. This is due to the fact that they share the same quantum circuit framework. Compared with version I (Fig. 1(a)), the computational times of quantum FS-LBM-II and quantum LKS-II are reduced to approximately one-third in 2D and one-quarter in 3D cases, respectively. This improvement arises because, for incompressible isothermal flows, quantum FS-LBM-II and quantum LKS-II require only one quantum circuit, while quantum FS-LBM-I and quantum LKS-I require three or four circuits. Therefore, quantum FS-LBM-II can effectively improve computational efficiency.

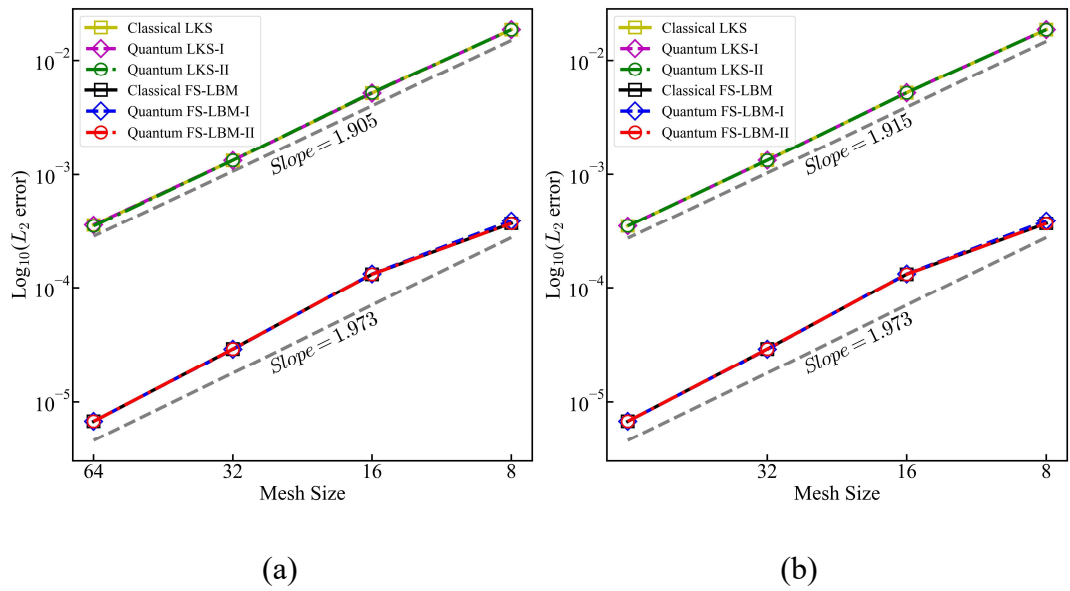

(a)          (b)

**Fig. 3** Convergence order of the $L_2$-norm versus mesh size for (a) 2D and (b) 3D Taylor-Green vortex flows.

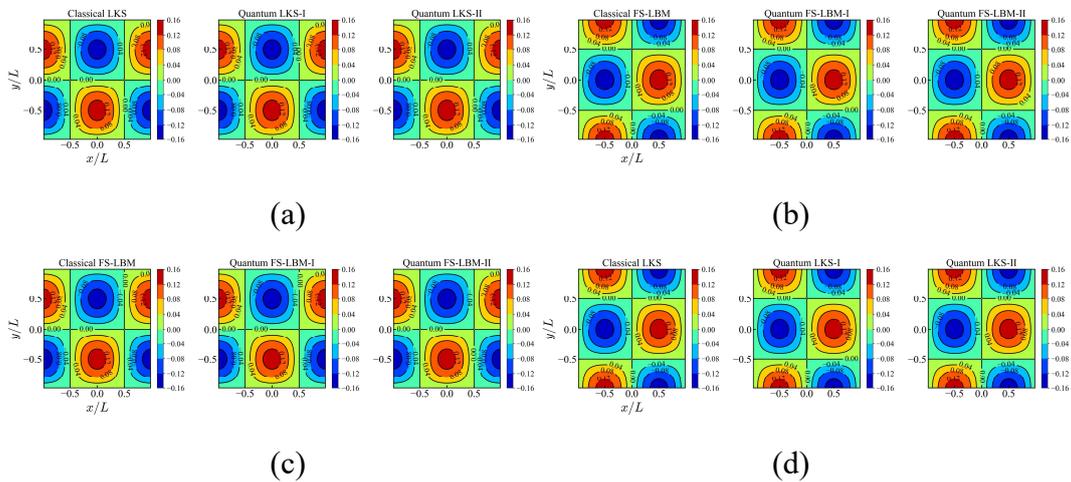

(a)          (b)

(c)          (d)

**Fig. 4** Comparison of (a) *u*- and (b) *v*-velocity contours for the 2D Taylor-Green vortex flow obtained by classical LKS, quantum LKS-I, and quantum LKS-II with a mesh size of 64 × 64. Comparison of (c) *u*- and (d) *v*-velocity contours for the 2D Taylor-Green vortex flow obtained by classical FS-LBM, quantum FS-LBM-I, and quantum FS-LBM-II with a mesh size of 64 × 64.

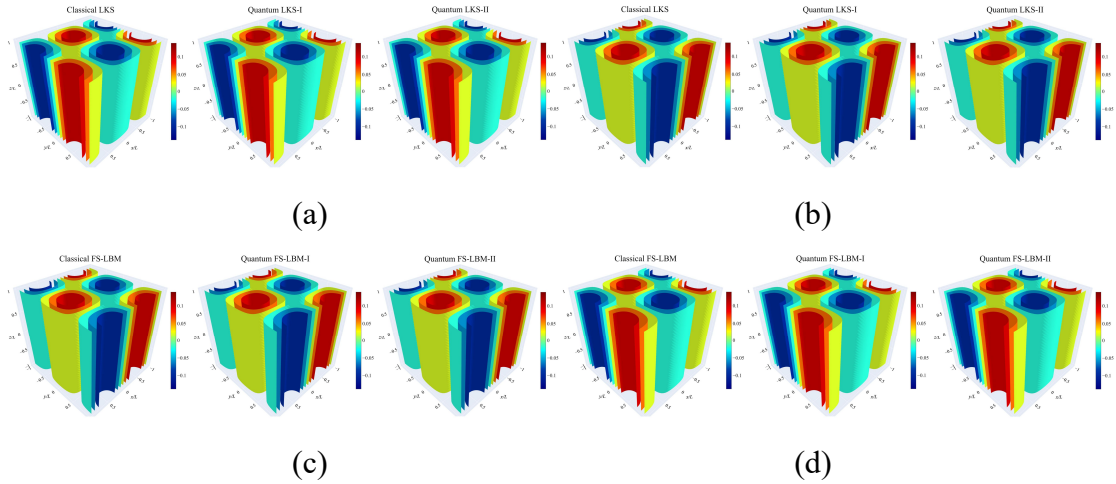

(a)　　　　　　　　　　　　　　(b)

(c)　　　　　　　　　　　　　　(d)

**Fig. 5** Comparison of (a) *u*- and (b) *v*-velocity contours for the 3D Taylor-Green vortex flow obtained by classical LKS, quantum LKS-I, and quantum LKS-II with a mesh size of 64 × 64 × 64. Comparison of (c) *u*- and (d) *v*-velocity contours for the 3D Taylor-Green vortex flow obtained by classical FS-LBM, quantum FS-LBM-I, and quantum FS-LBM-II with a mesh size of 64 × 64 × 64.

**Table 1** Computational times of quantum FS-LBM-I, quantum FS-LBM-II, quantum LKS-I, and quantum LKS-II for solving 2D and 3D Taylor-Green vortex flows at different mesh sizes.

| Methods | 2D (*s*) | 3D (*s*) |
| --- | --- | --- |

|                  | $N$=8 | $N$=16 | $N$=32 | $N$=64 | $N$=8 | $N$=16 | $N$=32 | $N$=64 |
|------------------|-------|--------|--------|--------|-------|--------|--------|--------|
| Quantum LKS-I    | 2.1   | 17.9   | 261.0  | 3605.2 | 26.8  | 613.7  | 17532.8| 560687.7|
| Quantum LKS-II   | 0.7   | 5.8    | 86.2   | 1181.3 | 6.3   | 144.9  | 4098.2 | 134245.3|
| Quantum FS-LBM-I | 2.1   | 18.2   | 261.7  | 3594.4 | 26.7  | 625.9  | 17157.8| 566130.3|
| Quantum FS-LBM-II| 0.7   | 5.9    | 84.9   | 1151.9 | 6.4   | 147.2  | 4093.5 | 132264.2|

**4.2 Two- and three-dimensional lid-driven cavity flows**

To further assess robustness and practical applicability, we simulate the 2D and 3D lid-driven cavity flows, a classical benchmark problem for validating newly developed numerical methods. Dirichlet boundary conditions are imposed on all walls, with the top lid moving at a constant velocity of $u = 0.1$, while the remaining walls are stationary. Notably, the Dirichlet boundary conditions in this case are implemented on the classical computer after computing macroscopic quantities from the quantum circuit. This treatment differs from the periodic boundary conditions, which are directly implemented within the quantum circuit, as discussed in Section 4.1.

For the 2D case, simulations were conducted on three different uniform meshes ($N \times N$, $N = 16$, 32, and 64) at Reynolds numbers of 100, 400, and 1000, with parameters $\delta_t = 1$, $\Delta x = 1$, and $\rho_0 = 1$. The velocity profiles along the cavity centerlines obtained by classical FS-LBM, quantum FS-LBM-I, and quantum FS-LBM-II are compared with those of classical LKS, quantum LKS-I, quantum LKS-II, as well as the benchmark results of Ghia et al. [44], as depicted in Figs. 6(a)-(c) for $Re = 100$, 400, and 1000, respectively. The results of the classical and quantum FS-LBMs show perfect consistency, further confirming the accuracy of the quantum FS-LBMs. In contrast, although the classical and quantum LKSs yield mutually consistent results, their accuracy is significantly lower than that of the classical and quantum FS-LBMs, particularly on coarse meshes and at high Reynolds numbers. Figs. 7(a)–(c) further compare convergence histories of the classical and quantum FS-LBMs as well as the classical and quantum LKSs under different mesh sizes and Reynolds numbers. Notably, convergence is achieved and the simulation is terminated once the residual drops below $10^{-6}$. The classical and quantum FS-LBMs exhibit almost identical convergence trends with comparable iteration counts, all consistently lower than those of the classical and quantum LKSs. These results clearly demonstrate that the quantum FS-LBMs offer superior computational accuracy and efficiency over the quantum LKSs.

To further verify the applicability of the quantum FS-LBMs at higher Reynolds number flows, we simulated the 2D lid-driven cavity flow at $Re = 5000$ on a mesh of $64 \times 64$. Figs. 8(a) and 8(b) present the velocity profiles along the vertical and horizontal centerlines, together with the convergence histories, compared with results from the classical FS-LBM, quantum FS-LBM-I, quantum FS-LBM-II, and the benchmark data of Ghia et al. [44]. The corresponding flow patterns obtained from the classical and quantum FS-LBMs are shown in Fig. 8(c). Overall, the classical and quantum FS-LBMs yield generally consistent results, successfully capturing the secondary vortices at the cavity bottom and the upper-left corner. However, while the convergence histories of the classical FS-LBM and quantum FS-LBM-II are identical, quantum FS-LBM-I shows slight deviations, suggesting that the quantum circuit used in quantum FS-LBM-I for computing macroscopic quantities may introduce additional errors. A similar behavior is also observed for the classical and quantum LKSs in Fig. 8(b). Although none achieves convergence, the quantum LKS-I diverges earlier than both the classical LKS and quantum LKS-II. This further highlights the superior robustness of quantum FS-LBMs compared to quantum LKSs. In addition, since the results of the quantum FS-LBMs become more accurate with increasing mesh size, we simulated the 2D lid-driven cavity flow at $Re = 1000$ using a mesh of 128 × 128. Figs. 9(a) and 9(b) present the velocity profiles along the vertical and horizontal centerlines as well as the convergence histories obtained from the classical and quantum FS-LBMs as well as the classical and quantum LKSs, while Fig. 9(c) shows the corresponding flow patterns. At this resolution, the quantum LKS-I again fails to converge, while the classical and quantum FS-LBMs achieve convergence with superior accuracy and efficiency compared to the classical LKS and quantum LKS-II. These findings further confirm the greater robustness of the quantum FS-LBMs over the quantum LKSs.

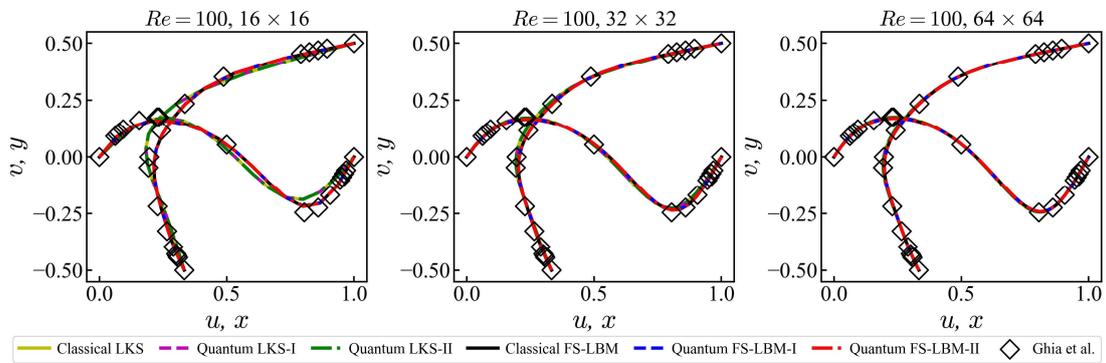

(a) $Re = 100$

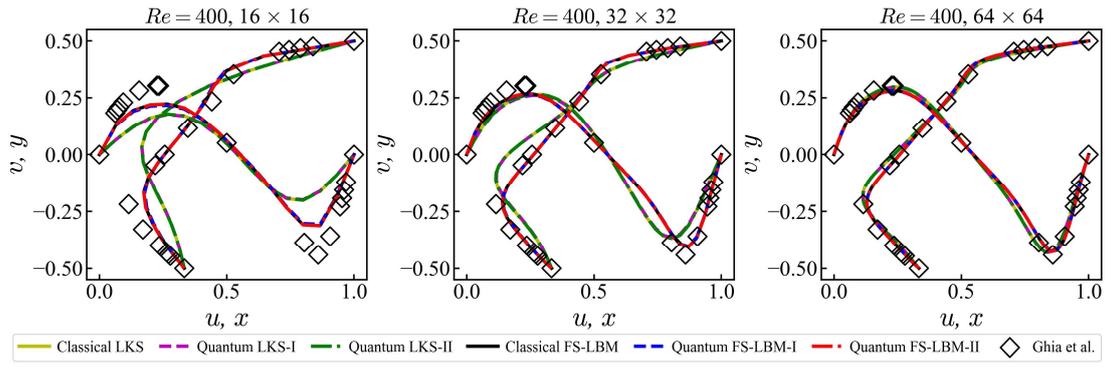

(b) *Re* = 400

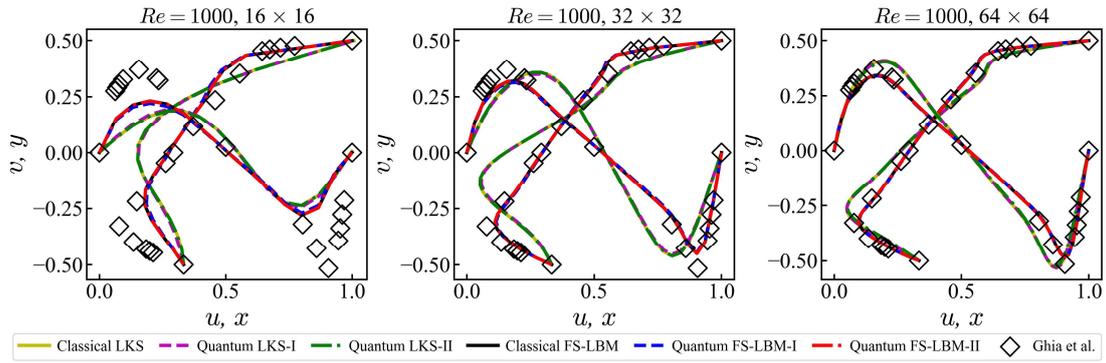

(c) *Re* = 1000

**Fig. 6** Comparison of vertical and horizontal centerline velocity profiles with the benchmark results of Ghia et al. [44] at (a) *Re* = 100, (b) *Re* = 400, and (c) *Re* = 1000.

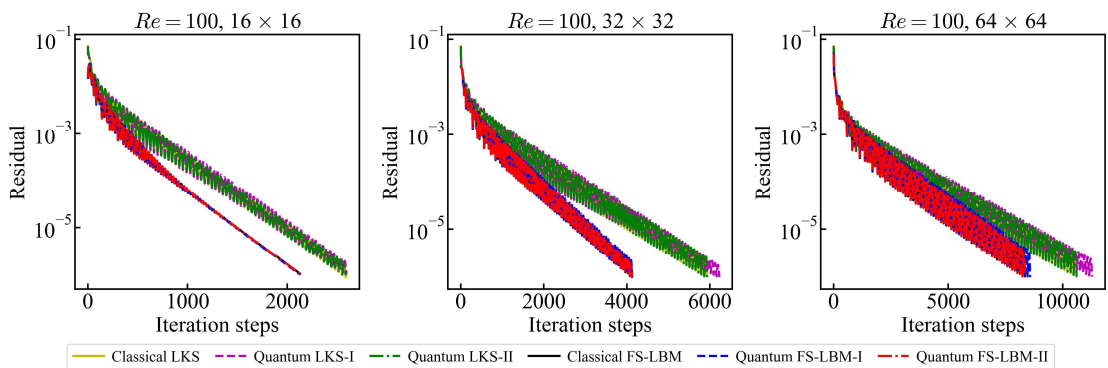

(a) *Re* = 100

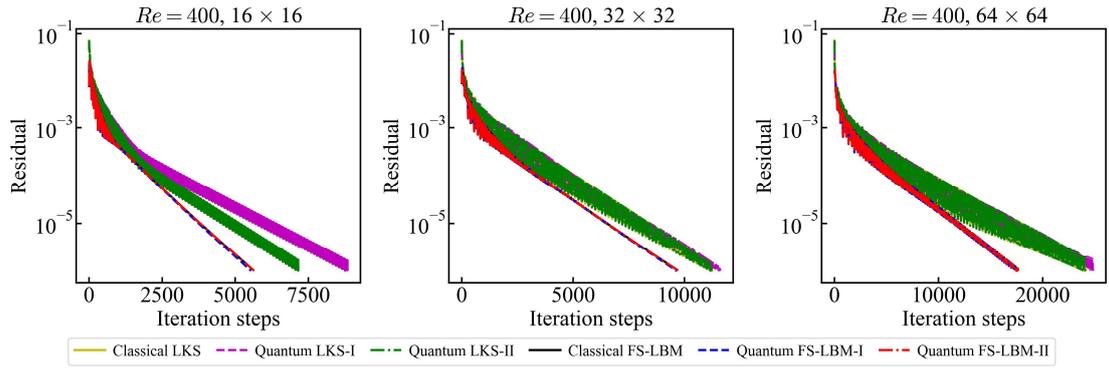

(b) *Re* = 400

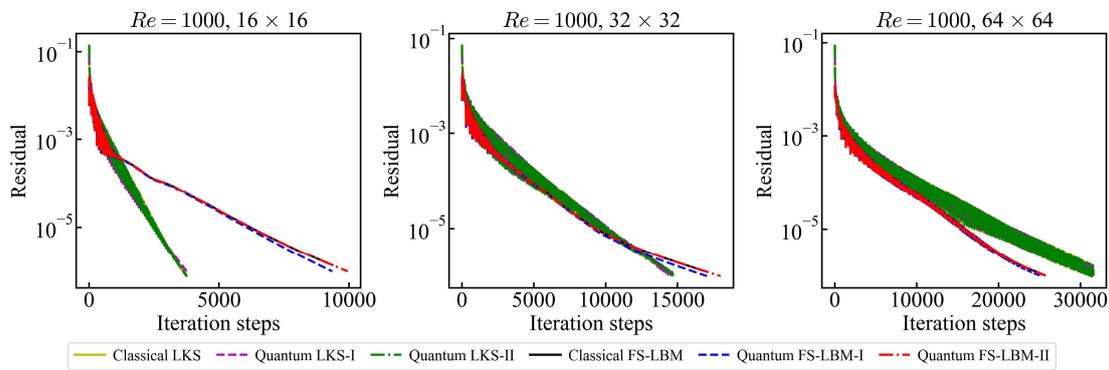

(c) *Re* = 1000

**Fig. 7** Convergence history of the classical FS-LBM, quantum FS-LBM-I, quantum FS-LBM-II, classical LKS, quantum LKS-I, and quantum LKS-II with different mesh sizes at (d) *Re* = 100, (e) *Re* = 400, and (f) *Re* = 1000.

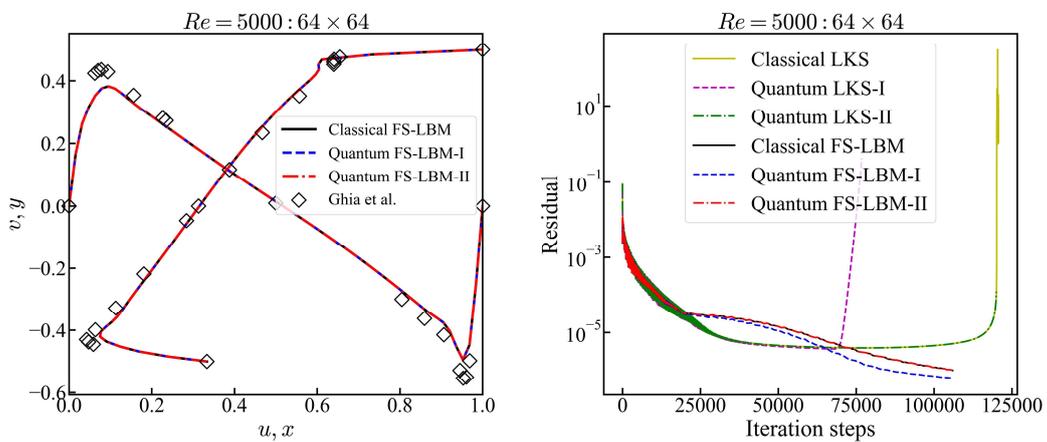

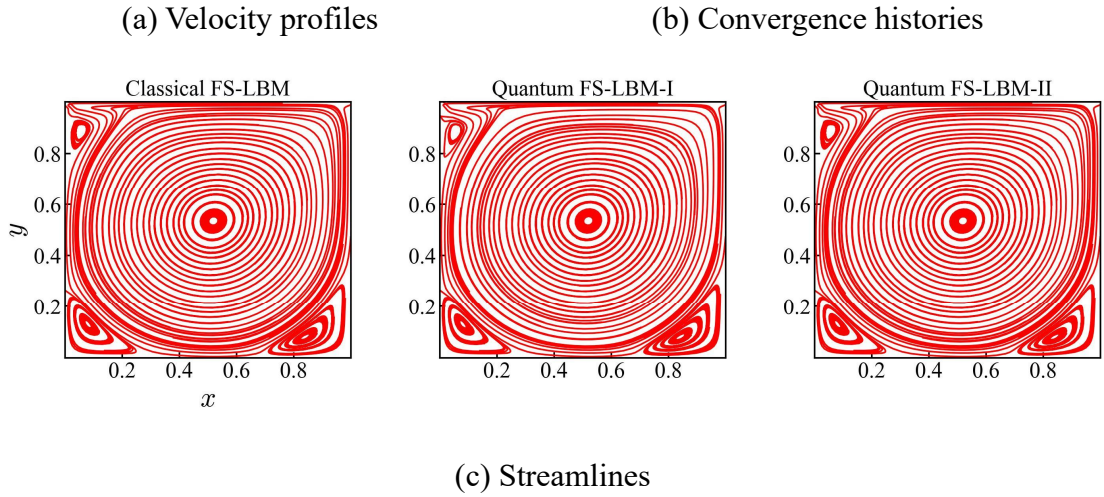

(c) Streamlines

**Fig. 8** Comparison of (a) vertical and horizontal centerline velocity profiles and (b) convergence histories obtained by classical FS-LBM, quantum FS-LBM-I, quantum FS-LBM-II, classical LKS, quantum LKS-I, and quantum LKS-II at $Re$ = 5000. (c) Streamlines of the 2D lid-driven cavity flow at $Re$ = 5000 obtained by the same methods.

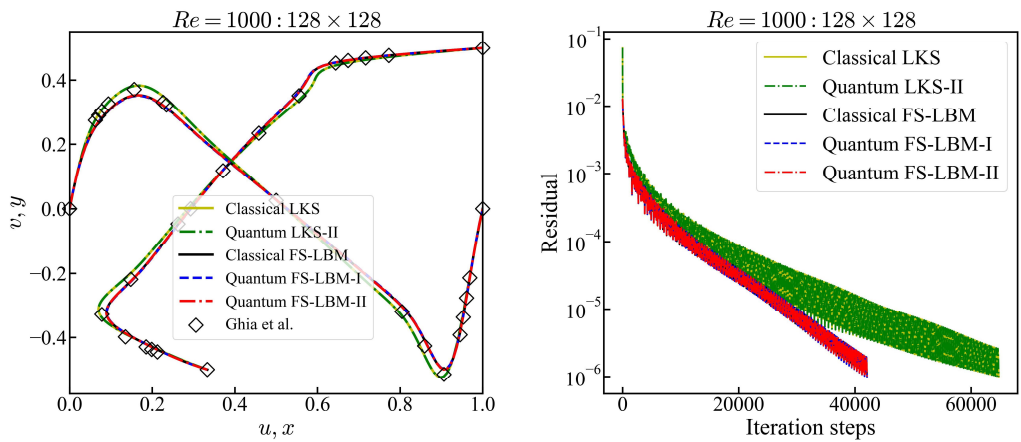

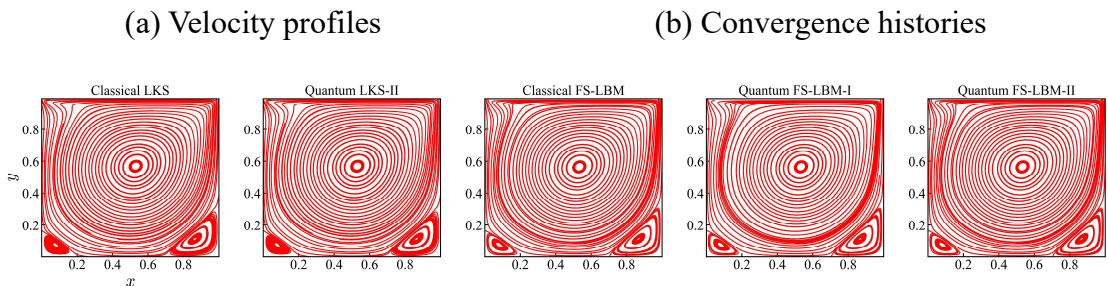

(c) Streamlines

**Fig. 9** Comparison of (a) vertical and horizontal centerline velocity profiles and (b) convergence histories obtained by classical FS-LBM, quantum FS-LBM-I, quantum FS-LBM-II, classical LKS, quantum LKS-I, and quantum LKS-II at $Re$ = 1000. (c) Streamlines of the 2D lid-driven cavity flow at $Re$ = 1000 obtained by the same methods.

Finally, we performed a simulation of the 3D cavity flow at $Re$ = 100 on a mesh of 32 × 32 × 32. Notably, the Laplacian operator in the corrector step of Eq. (8) was discretized using the SS stencil described in Eq. (10). Figs. 10(a)–(c) depict the velocity vector fields of the classical and quantum LKSs on the $x$–$z$, $y$–$z$, and $x$–$y$ planes at the center of the cavity, whereas Figs. 10(d)–(f) show the corresponding velocity vector fields for the classical and quantum FS-LBMs. It can be seen that both methods exhibit strong consistency between their classical and quantum implementations, with similar overall flow patterns. Moreover, Figs. 11 and 12 present the $u$-velocity distribution along the vertical centerline of the cavity and the corresponding convergence histories for the classical and quantum FS-LBMs and LKSs, respectively. From Fig. 11, all results are in close agreement with the benchmark data of Wong and Baker [45] and Jiang et al. [46]. While slight deviations appear in the central region of the velocity profiles, the classical and quantum FS-LBMs provide more accurate predictions in the upper-right region compared to their LKS counterparts. Additionally, the classical and quantum FS-LBMs converge with fewer iterations than the classical and quantum LKSs. These results demonstrate the effectiveness and robustness of quantum FS-LBMs in simulating complex three-dimensional incompressible flows.

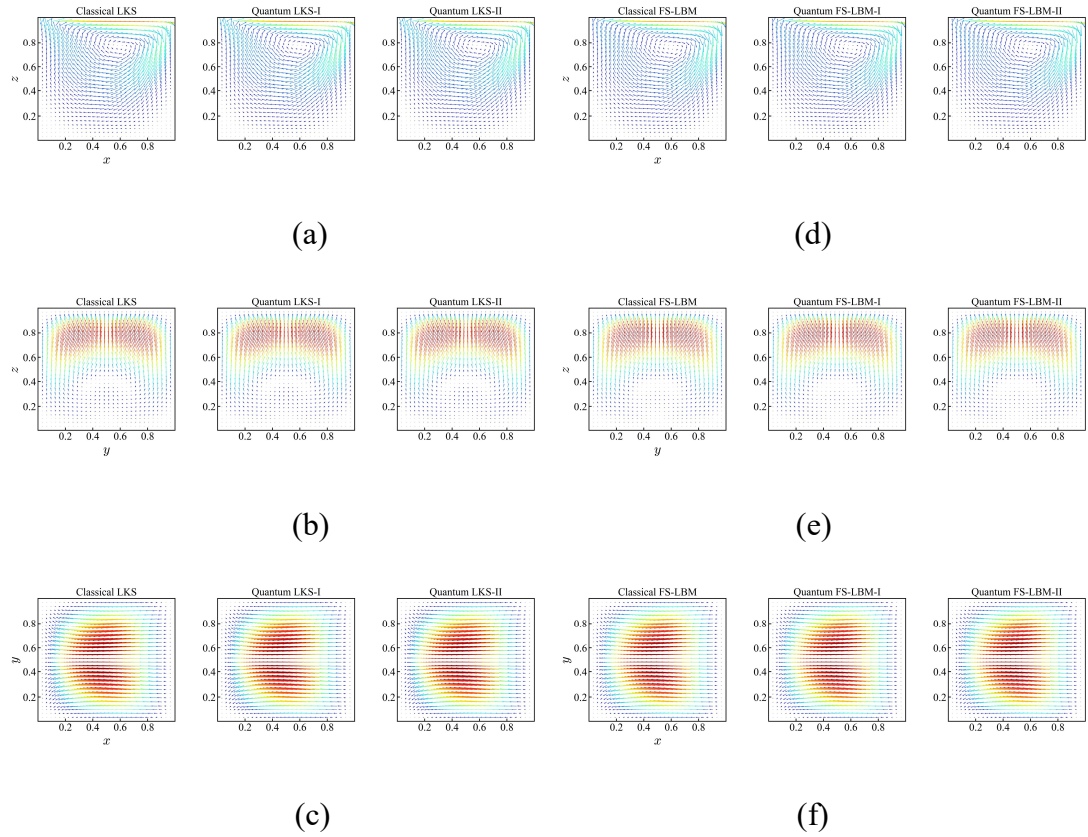

**Fig. 10** Flow patterns in vector form on the centroidal planes of (a) *x-z*, (b) *y-z*, and (c) *x-y* obtained by classical LKS, quantum LKS-I, and quantum LKS-II. Flow patterns in vector form on the centroidal planes of (d) *x-z*, (e) *y-z*, and (f) *x-y* obtained by classical FS-LBM, quantum FS-LBM-I, and quantum FS-LBM-II.

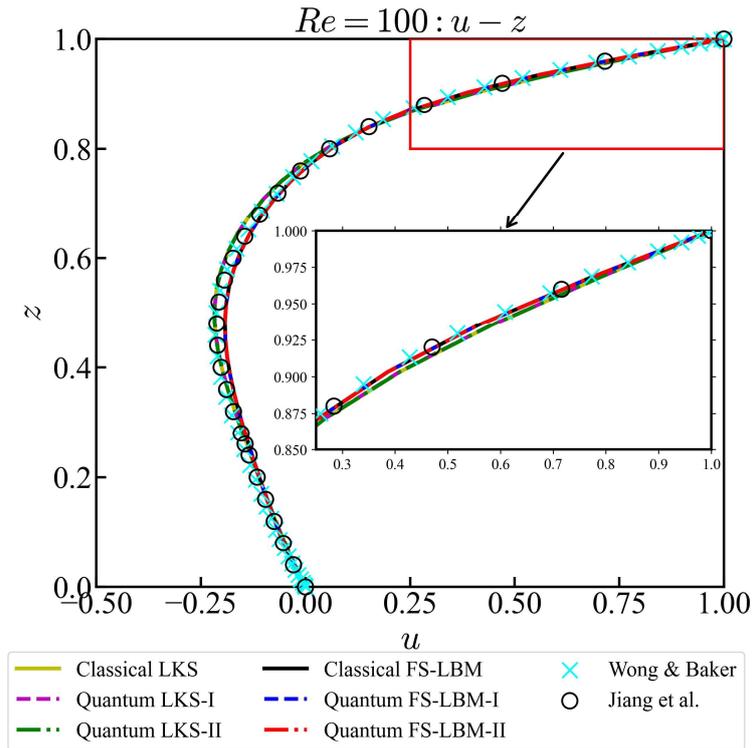

**Fig. 11** Comparison of the *u*-velocity component distribution along the vertical centerline with benchmark results from Wong and Baker [45] and Jiang et al. [46].

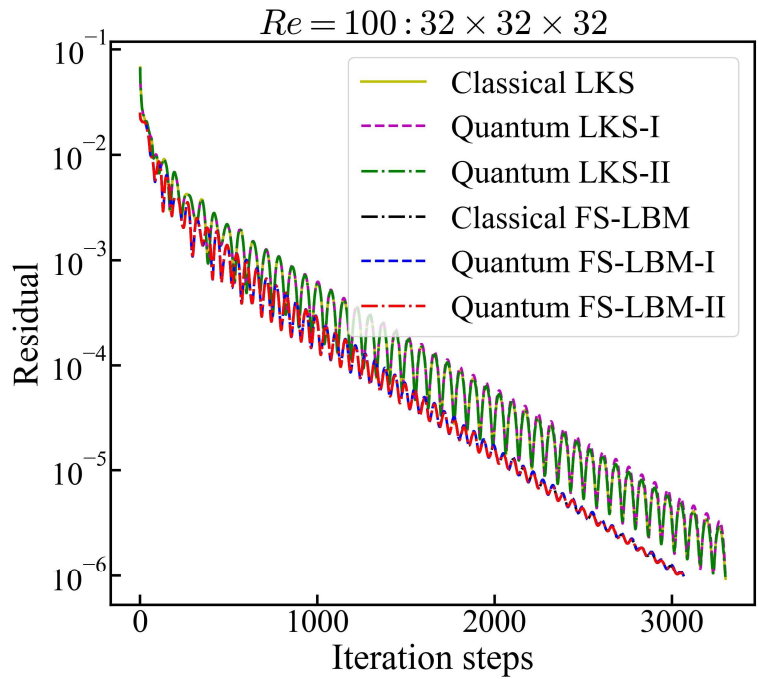

**Fig. 12** Convergence history of classical FS-LBM, quantum FS-LBM-I, quantum

**4.3 Two- and three-dimensional natural convention in a cavity**

In this section, we validate the applicability of quantum FS-LBMs for incompressible thermal flows by simulating 2D and 3D natural convection in a square cavity. All walls are stationary and subject to Dirichlet boundary conditions. In the 2D case, the top and bottom walls are adiabatic, while the left and right walls are isothermal with $T_1 = 2$ and $T_0 = 1$. In the 3D case, the two walls perpendicular to the *x-z* plane are isothermal with $T_1 = 2$ and $T_0 = 1$, and the remaining walls are adiabatic. The dynamic similarity of this test case is characterized by two dimensionless parameters: the Prandtl number (*Pr*) and the Rayleigh number (*Ra*), which are defined as follows:

$$Pr = \frac{\upsilon}{\kappa}, \quad Ra = \frac{g\beta\Delta T H^3}{\upsilon\kappa}, \tag{23}$$

where $\Delta T = T_1 - T_0$, and $H$ is the length of the square cavity. In the simulation, the Nusselt number *Nu* is used to evaluate the heat transfer rate, and the average Nusselt number $\overline{Nu}$ over the entire computational domain is defined as

$$\overline{Nu} = \frac{H}{\kappa\Delta T}\frac{1}{H^2}\oint_\Omega \left(uT - \kappa\frac{\partial T}{\partial x}\right)d\Omega. \tag{24}$$

First, the 2D natural convection in a square cavity was simulated on three different uniform meshes ($N \times N$, $N = 16$, 32, and 64) at Rayleigh numbers of $10^3$, $10^4$, and $10^5$. The parameters are set to $\delta_t = 1$, $\Delta x = 1$, $Pr = 0.71$, and $g\beta = 10^{-5}$.

Convergence is achieved once the residual drops below $10^{-7}$. Figs. 13(b) and 13(d) present the isotherms obtained by the classical and quantum FS-LBMs at $Ra = 10^3$ and $Ra = 10^4$, while the corresponding results from the classical and quantum LKSs are shown in Figs. 13(a) and 13(c). It can be observed that the temperature contours from the classical and quantum FS-LBMs coincide perfectly under the same mesh size, demonstrating the effectiveness of quantum FS-LBMs in simulating incompressible thermal flows. Compared with the results of the classical and quantum LKSs, which require a mesh size of 64 × 64 to obtain reasonable accuracy, the classical and quantum FS-LBMs deliver superior accuracy and attain reliable results on a coarser mesh size of 32 × 32. Figs. 14(a) and 14(b) show the convergence histories of the classical and quantum FS-LBMs as well as the classical and quantum LKSs at $Ra = 10^3$ and $Ra = 10^4$, respectively. The convergence curves of the classical and quantum FS-LBMs coincide exactly, but their iteration counts are higher than those of the classical and quantum LKSs, especially on coarse meshes. This discrepancy is likely because the classical and quantum LKSs converge to inaccurate solutions, as evidenced by the isotherms in Figs. 13(a) and 13(c). Under coarse mesh conditions, the temperature contours produced by the classical and quantum LKSs lose centro-symmetry, while those from the classical and quantum FS-LBMs remain fully centro-symmetric across all mesh sizes.

For $Ra = 10^5$, Fig. 15(a) depicts the isotherms obtained by the classical and quantum FS-LBMs, while Fig. 15(b) illustrates the corresponding convergence histories alongside those of the classical and quantum LKSs. Although the classical and quantum FS-LBMs again produce highly consistent results at the same mesh size, reasonable solutions are achieved only on the mesh size of 64 × 64. In contrast, the classical and quantum LKSs diverge outright on the mesh size of 32 × 32. Even on the mesh size of 64 × 64, they fail to converge, with residuals oscillating around $10^{-3}$, yielding incorrect solutions as shown in Fig. 16(b). Moreover, while the classical and quantum LKSs appear to converge structurally on the coarse mesh size of 16 × 16, their results remain erroneous, as shown in Fig. 16(a). Table 2 presents a quantitative comparison of representative physical parameters obtained from the classical and quantum FS-LBMs and LKSs against the benchmark results of Wang et al. [47] and Yang [48]. Here, $u_{max}$ denotes the maximum $u$-velocity along the vertical centerline with $y$ indicating the corresponding coordinate, while $v_{max}$ denotes the maximum $v$-velocity along the horizontal centerline with $x$ indicating the corresponding coordinate. It can be observed that the results of the classical and quantum FS-LBMs on the mesh size of 64×64 exhibit excellent consistency with the reference data and demonstrate superior accuracy compared to the classical and quantum LKSs.

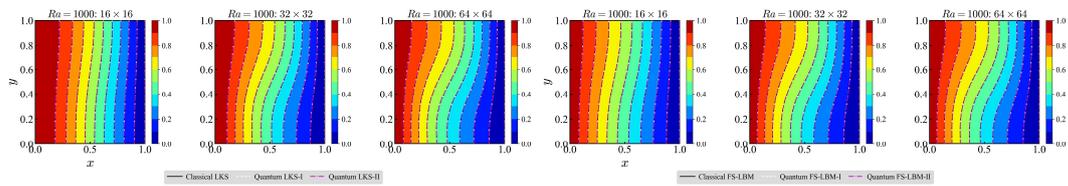

(a)                                    (b)

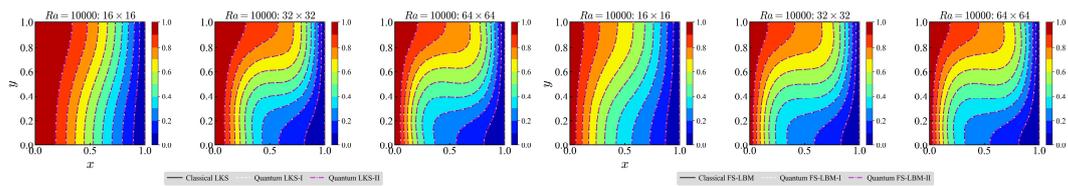

(c)                                    (d)

**Fig. 13** Isotherms obtained by the classical LKS, quantum LKS-I, and quantum LKS-II at (a) $Ra = 10^3$ and (c) $Ra = 10^4$. Isotherms obtained by the classical FS-LBM, quantum FS-LBM-I, and quantum FS-LBM-II at (b) $Ra = 10^3$, (d) $Ra = 10^4$, and (f) $Ra = 10^5$.

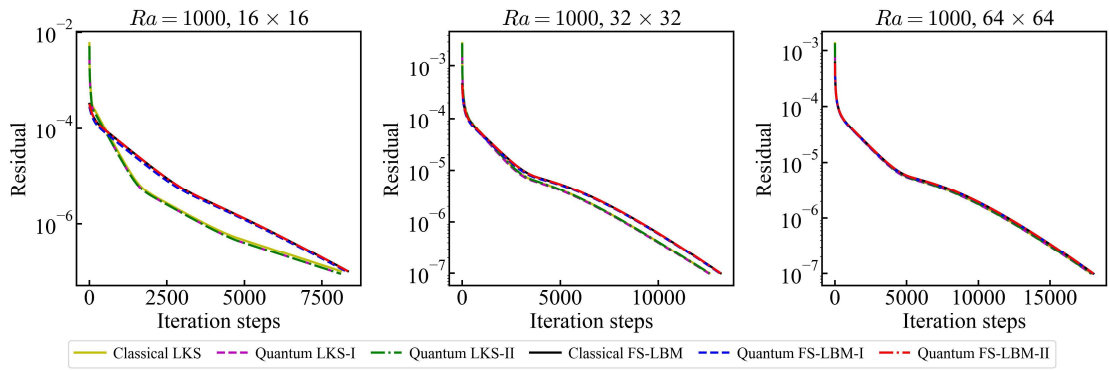

(a) $Ra = 1000$

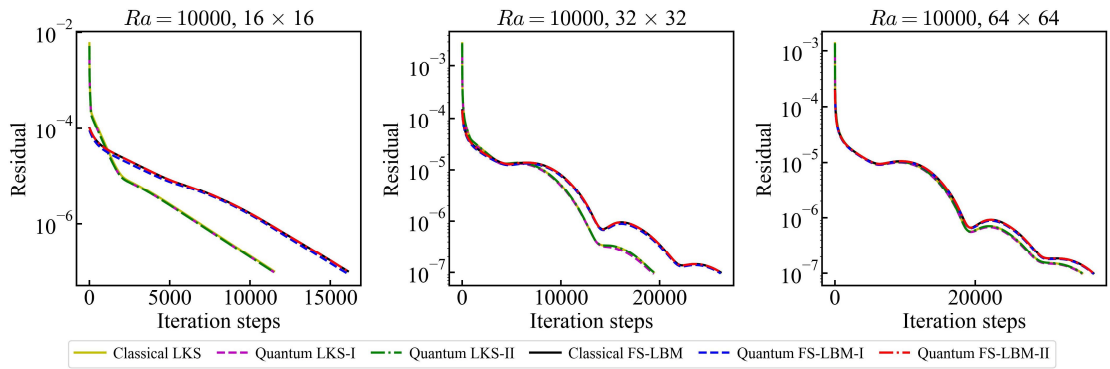

(b) $Ra = 10000$

**Fig. 14** Convergence histories of classical FS-LBM, quantum FS-LBM-I, quantum FS-LBM-II, classical LKS, quantum LKS-I, and quantum LKS-II with different mesh sizes for the 2D natural convection in a square cavity at (a) $Ra = 10^3$ and (b) $Ra = 10^4$

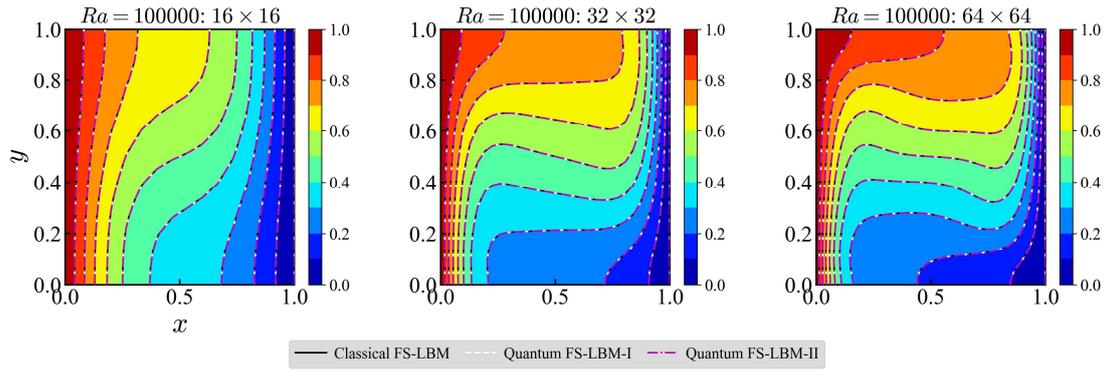

(a) Isotherms

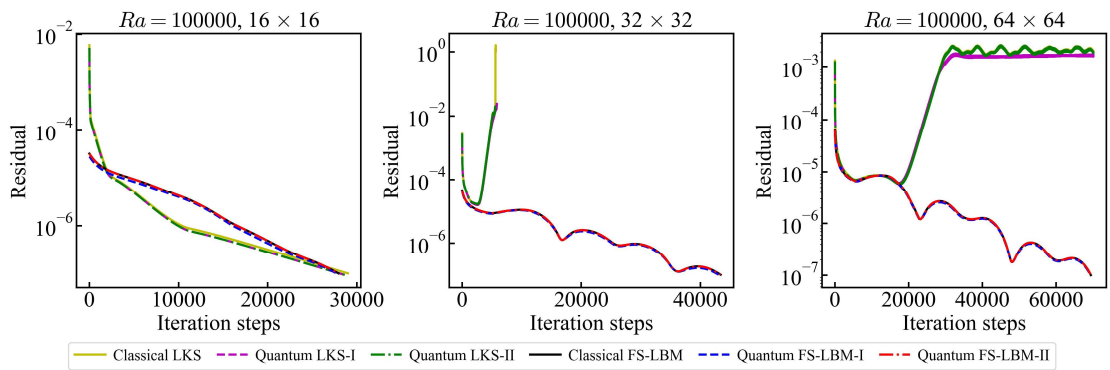

(b) Convergence histories

**Fig. 15** (a) Isotherms obtained by the classical FS-LBM, quantum FS-LBM-I, and quantum FS-LBM-II at $Ra = 10^5$. (b) Convergence histories of classical FS-LBM, quantum FS-LBM-I, quantum FS-LBM-II, classical LKS, quantum LKS-I, and quantum LKS-II with different mesh sizes for the 2D natural convection in a square cavity at $Ra = 10^5$.

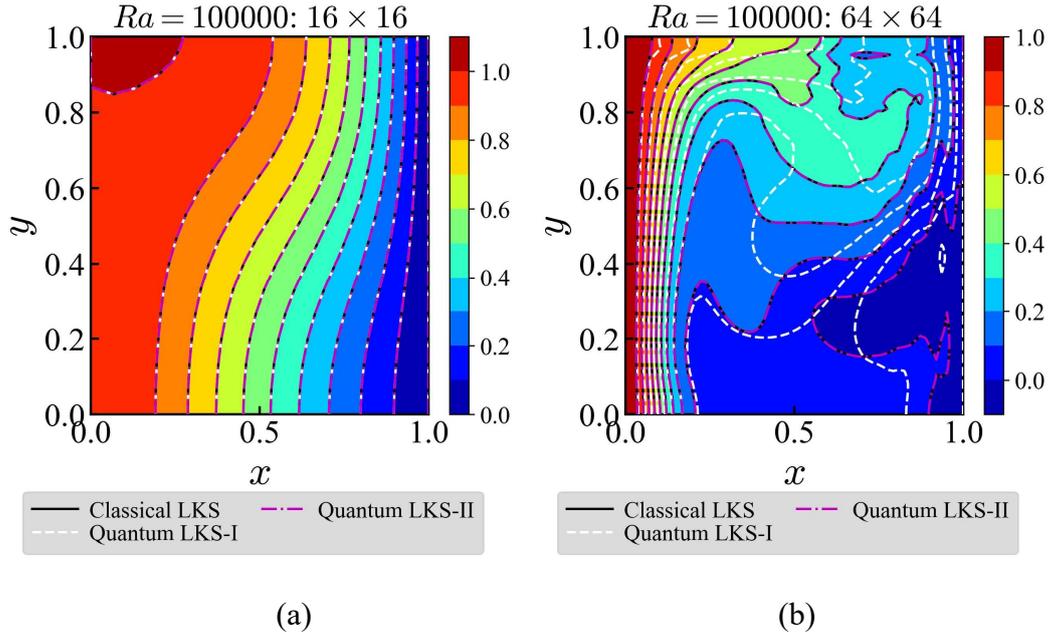

(a)  (b)

**Fig. 16** Isotherms obtained by the classical LKS, quantum LKS-I, and quantum LKS-II with the mesh sizes of (a) 16 × 16 and (b) 64 × 64 at $Ra = 10^5$.

**Table 2** Results of natural convection in a square cavity at different Rayleigh numbers of $Ra = 10^3$, $10^4$, and $10^5$.

| $Ra$ | | Classical LKS | Quantum LKS-I | Quantum LKS-II | Classical FS-LBM | Quantum FS-LBM-I | Quantum FS-LBM-II | Wang et al. | Yang et al. |
|---|---|---|---|---|---|---|---|---|---|
| $10^3$ (2D) | $u_{max}$ | 3.561 | 3.561 | 3.561 | **3.639** | **3.639** | **3.639** | 3.640 | 3.647 |
| | $y$ | 0.8095 | 0.8095 | 0.8095 | **0.8095** | **0.8095** | **0.8095** | 0.815 | 0.815 |
| | $v_{max}$ | 3.597 | 3.597 | 3.597 | **3.686** | **3.686** | **3.686** | 3.708 | 3.696 |
| | $x$ | 0.1905 | 0.1905 | 0.1905 | **0.1746** | **0.1746** | **0.1746** | 0.180 | 0.175 |
| | $\overline{Nu}$ | 1.121 | 1.121 | 1.121 | **1.114** | **1.114** | **1.115** | 1.115 | 1.118 |
| $10^4$ (2D) | $u_{max}$ | 16.306 | 16.305 | 16.306 | **16.124** | **16.123** | **16.124** | 16.140 | 16.183 |
| | $y$ | 0.8254 | 0.8254 | 0.8254 | **0.8254** | **0.8254** | **0.8254** | 0.825 | 0.825 |
| | $v_{max}$ | 18.832 | 18.832 | 18.832 | **19.438** | **19.438** | **19.438** | 19.670 | 19.627 |
| | $x$ | 0.1587 | 0.1587 | 0.1587 | **0.1269** | **0.1269** | **0.1269** | 0.118 | 0.117 |
| | $\overline{Nu}$ | 2.190 | 2.190 | 2.190 | **2.216** | **2.216** | **2.216** | 2.232 | 2.245 |
| $10^5$ (2D) | $u_{max}$ | 121.3713 | 134.83023 | 121.3713 | **34.6144** | **34.6093** | **34.6144** | 34.870 | 34.775 |
| | $y$ | 0.9048 | 0.9048 | 0.9048 | **0.8571** | **0.8571** | **0.8571** | 0.855 | 0.853 |
|   | $v_{max}$ | 94.7511 | 99.5179 | 94.7511 | **67.0643** | **67.0592** | **67.0643** | 68.850 | 68.634 |
|   | $x$ | 0.1111 | 0.1269 | 0.1111 | **0.0635** | **0.0635** | **0.0635** | 0.065 | 0.067 |
|   | $\overline{Nu}$ | 4.681 | 4.877 | 4.681 | **4.403** | **4.403** | **4.403** | 4.491 | 4.524 |
|   | $u_{max}$ | 0.1248 | 0.1238 | 0.1230 | **0.1291** | **0.1280** | **0.1280** | 0.132 | 0.132 |
| $10^3$ | $y$ | 0.1935 | 0.1935 | 0.1935 | **0.1935** | **0.1935** | **0.1935** | 0.187 | 0.179 |
| (3D) | $v_{max}$ | 0.1198 | 0.1235 | 0.1231 | **0.1262** | **0.1281** | **0.1281** | 0.133 | 0.133 |
|   | $x$ | 0.8065 | 0.8065 | 0.8065 | **0.8065** | **0.8065** | **0.8065** | 0.829 | 0.821 |
|   | $\overline{Nu}$ | 1.0184 | 1.0613 | 1.0676 | **1.0518** | **1.0802** | **1.0802** | 1.092 | 1.088 |

Next, we simulate the 3D natural convection in a square cavity at $Ra = 10^3$ and $Ra = 10^4$ using a uniform mesh size of $32 \times 32 \times 32$. Figs. 17(a) and 17(b) present the isotherms on the $z = 0.5H$ plane at $Ra = 10^3$, obtained by the classical and quantum LKSs as well as classical and quantum FS-LBMs. The isotherms from the classical and quantum FS-LBMs are highly consistent, while those from the classical and quantum LKSs are also mutually consistent. Fig. 17(c) shows the convergence histories of both FS-LBMs and LKSs, revealing that the classical and quantum FS-LBMs maintain closely matching convergence curves, with marginally higher iteration counts than the classical and quantum LKSs. Fig. 17(d) compares the $u$-velocity along the vertical centerline and the $v$-velocity along the horizontal centerline on the $z = 0.5H$ plane against the benchmark results of Yang et al. [49]. The FS-LBMs, both classical and quantum, clearly achieve higher accuracy than the LKSs. This observation is further supported by the quantitative comparison of representative physical parameters in Table 2, which shows that the FS-LBMs yield results in closer agreement with the benchmark data from Yang et al. [49] and Wang et al. [50]. Finally, we extend the simulation to $Ra = 10^4$ using the quantum FS-LBMs on a mesh size of $32 \times 32 \times 32$. Figs. 18(a) and 18(b) show the isotherms and the $u$-velocity along the vertical centerline as well as the *v*-velocity along the horizontal centerline on the $z = 0.5H$ plane computed by the classical and quantum FS-LBMs, respectively. Similarly, the results of the quantum FS-LBM-I and quantum FS-LBM-II are in excellent agreement with those of the classical FS-LBM and show good consistency with the results of Yang et al. [49], further confirming the applicability and robustness of quantum FS-LBMs in simulating incompressible thermal flows.

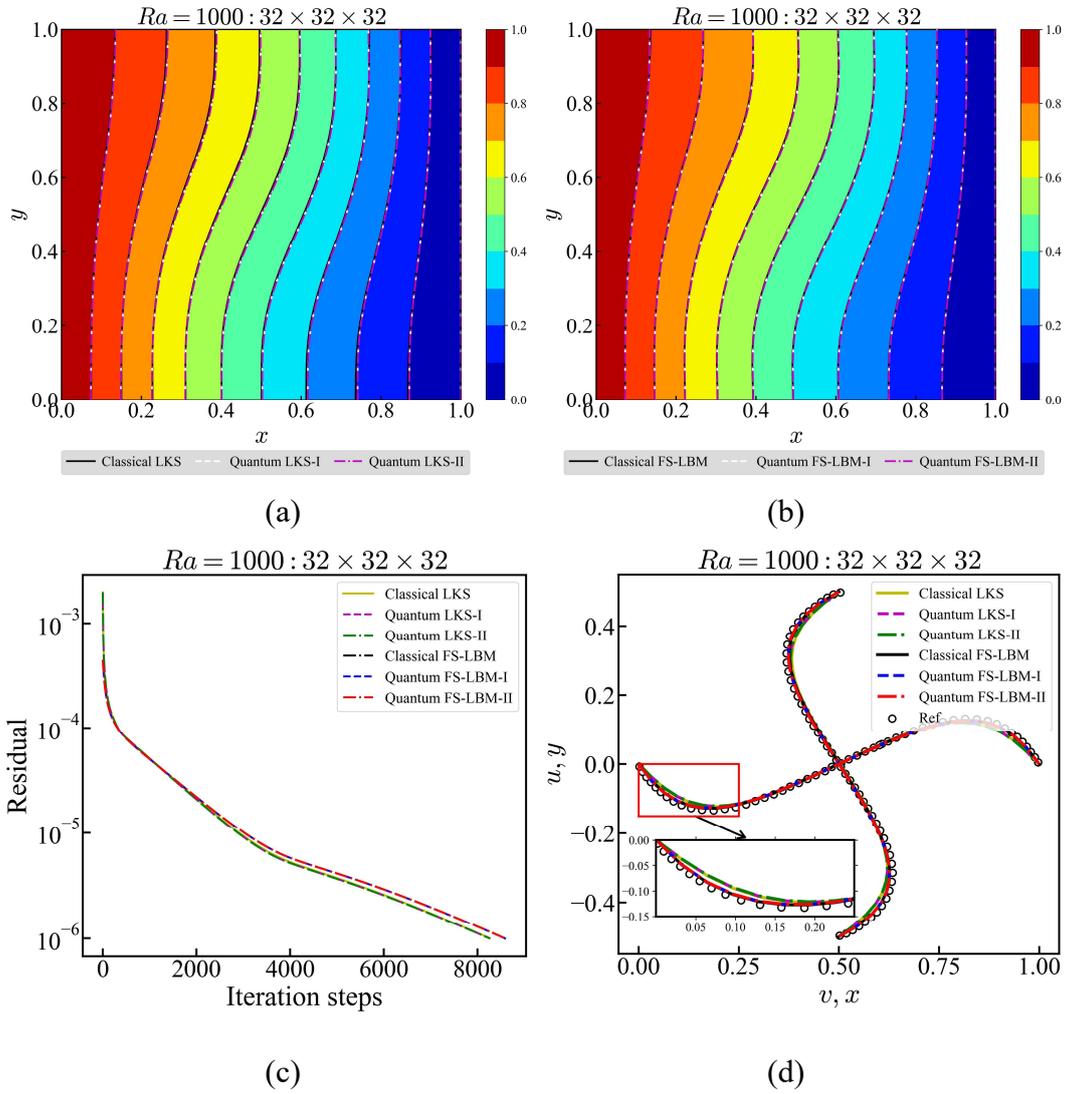

**Fig. 17** Isotherms on the $z = 0.5H$ plane at $Ra = 10^3$ obtained by (a) classical LKS, quantum LKS-I, quantum LKS-II, and (b) classical FS-LBM, quantum FS-LBM-I, quantum FS-LBM-II. (c) Convergence histories of classical FS-LBM, quantum

FS-LBM-I, quantum FS-LBM-II, classical LKS, quantum LKS-I, and quantum LKS-II at $Ra = 10^3$. (d) Comparison of the $u$-velocity along the vertical centerline and the $v$-velocity along the horizontal centerline on the $z = 0.5H$ plane obtained by classical LKS, quantum LKS-I, quantum LKS-II, classical FS-LBM, quantum FS-LBM-I, and quantum FS-LBM-II against the benchmark results from Yang et al. [49].

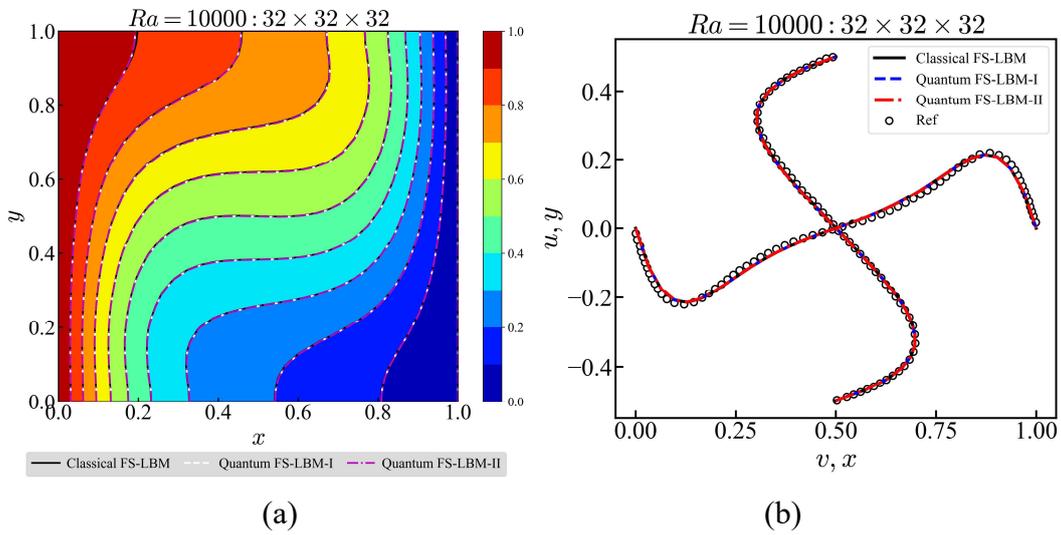

**Fig. 18** (a) Isotherms on the $z = 0.5H$ plane at $Ra = 10^4$ obtained by classical FS-LBM, quantum FS-LBM-I, and quantum FS-LBM-II. (b) Comparison of the $u$-velocity along the vertical centerline and the $v$-velocity along the horizontal centerline on the $z = 0.5H$ plane obtained by classical FS-LBM, quantum FS-LBM-I, and quantum FS-LBM-II against the results from Yang et al. [49].

## 5 Conclusions

In this study, we proposed a quantum FS-LBM for simulating two- and three-dimensional incompressible isothermal and thermal flows at arbitrary Reynolds and Rayleigh numbers. The method comprises a predictor step and a corrector step: the predictor is performed via the LBM on a quantum circuit, while the corrector is computed using finite differences on a classical computer. Additionally, computing the density and velocity fields separately would require multiple quantum circuits, consuming substantial quantum resources. To overcome this, we developed a quantum FS-LBM-II variant that employs a single quantum circuit to generate the post-streaming distribution functions, from which both fields can then be computed on a classical computer. This design substantially reduces quantum resource requirements while enhancing computational efficiency.

The performance of the proposed method was evaluated via a series of benchmark tests on representative 2D and 3D incompressible isothermal and thermal flows. Numerical results indicate that both quantum FS-LBM-I and FS-LBM-II yield solutions highly consistent with the classical LKS at the same grid resolution and share the same convergence order. Moreover, the convergence trend of quantum FS-LBM-II aligns perfectly with that of the classical FS-LBM, whereas quantum FS-LBM-I shows slight deviations. This discrepancy may be attributed to additional errors introduced by the quantum circuits used for computing macroscopic quantities. Compared with quantum LKSs, the quantum FS-LBMs exhibit higher accuracy, stability, and efficiency.

In summary, the quantum FS-LBMs offer a practical and efficient framework for simulating incompressible flows on quantum hardware, potentially advancing large-scale fluid dynamics simulations. With the relaxation parameter set to 1, they can be seamlessly integrated into most existing quantum LBMs, demonstrating excellent scalability and wide applicability. Future work will focus on implementing finite-difference methods within quantum circuits, enabling the corrector step to be performed entirely on quantum hardware. Additionally, since any boundary condition in LBM can be expressed in the form of $A \cdot f_\alpha$, the matrix $A$ may be decomposed via singular value decomposition (SVD), with the resulting factors encoded into quantum circuits. This would allow non-periodic boundary conditions to be fully integrated into the quantum framework.

**Appendix A: Supplementary details of the quantum lattice kinetic scheme for incompressible thermal flows**

Similar to FS-LBM, LKS also sets the relaxation parameter to 1, so that its lattice Boltzmann equation for simulating incompressible thermal flows is given by

$$f_\alpha(\bm{x}+\bm{e}_\alpha \delta t, t+\delta t) = f_\alpha^{eq}(\bm{x},t), \\ h_\alpha(\bm{x}+\bm{e}_\alpha \delta t, t+\delta t) = h_\alpha^{eq}(\bm{x},t).$$
(A.1)

As discussed in Section 2, the N-S equations recovered from Eq. (A.1) yield both viscosity $\upsilon$ and thermal diffusivity $\kappa$ fixed at 1/6, meaning that a given mesh size can only simulate flow at a specific Reynolds number, which greatly restricts flexibility. To overcome this limitation, Inamuro [19] modified the EDF $f_\alpha^{eq}(\bm{x},t)$ as follows

$$f_\alpha^{eq}(\bm{x},t) = w_\alpha \rho \left( 1 + \frac{\bm{e}_\alpha \cdot \bm{u}}{c_s^2} + \frac{(\bm{e}_\alpha \cdot \bm{u})^2}{2c_s^4} - \frac{\bm{u} \cdot \bm{u}}{2c_s^2} + A \delta t \bm{e}_\alpha^T \left( \nabla \bm{u} + (\nabla \bm{u})^T \right) \bm{e}_\alpha \right),$$
(A.2)

where $A$ is a constant. Through the C-E expansion analysis [40], the viscosity $\upsilon$ is then related to the constant $A$ by

$$\upsilon = \left(\frac{1}{2} - 2Ac_s^2\right)c_s^2 \delta t. \tag{A.3}$$

Thus, the viscosity $\upsilon$ can be adjusted by varying $A$ instead of being fixed, allowing simulations at arbitrary Reynolds numbers. For the temperature equilibrium distribution function $h_\alpha^{eq}(\boldsymbol{x},t)$, we follow the method of Meng et al. [51] to modify it as

$$h_\alpha^{eq}(\boldsymbol{x},t) = w_\alpha T\left(1 + \frac{\boldsymbol{e}_\alpha \cdot \boldsymbol{u}}{c_s^2} + \frac{(\boldsymbol{e}_\alpha \cdot \boldsymbol{u})^2}{2c_s^4} - \frac{\boldsymbol{u} \cdot \boldsymbol{u}}{2c_s^2}\right) + w_\alpha B \delta t (\boldsymbol{e}_\alpha \cdot \nabla T), \tag{A.4}$$

Through the C-E expansion analysis [51], the thermal diffusivity $\kappa$ is then related to the constant $B$ by

$$\kappa = \left(\frac{1}{2} - B\right)c_s^2 \delta t. \tag{A.5}$$

Hence, the thermal diffusivity $\kappa$ can be tuned by adjusting $B$. By replacing the original EDFs with those defined in Eqs. (A.2) and (A.4), LKS can overcome the constraints of mesh size and enable simulations of flows at arbitrary Reynolds numbers.

For the implementation of quantum LKS, the quantum circuit framework and computational procedure are identical to those of the quantum FS-LBM. The only modification is that the collision matrix $D$ in Eq. (13) is replaced with the following form

$$D = \begin{bmatrix} C_0 D_0 & 0 & 0 \\ 0 & \ddots & 0 \\ 0 & 0 & C_{Q-1} D_{Q-1} \end{bmatrix}, \quad D_\alpha = \begin{bmatrix} f_\alpha^{eq}(\boldsymbol{x}_0,t) & 0 & 0 \\ 0 & \ddots & 0 \\ 0 & 0 & f_\alpha^{eq}(\boldsymbol{x}_{M-1},t) \end{bmatrix}, \tag{A.6}$$

where

$$f_\alpha^{eq}(\boldsymbol{x},t) = f_\alpha^{eq}(\boldsymbol{x},t)/\rho = w_\alpha\left(1 + \frac{\boldsymbol{e}_\alpha \cdot \boldsymbol{u}}{c_s^2} + \frac{(\boldsymbol{e}_\alpha \cdot \boldsymbol{u})^2}{2c_s^4} - \frac{\boldsymbol{u} \cdot \boldsymbol{u}}{2c_s^2} + A\delta t \boldsymbol{e}_\alpha^T\left(\nabla \boldsymbol{u} + (\nabla \boldsymbol{u})^T\right)\boldsymbol{e}_\alpha\right). \tag{A.7}$$

For the temperature equilibrium distribution function $h_\alpha^{eq}(\boldsymbol{x},t)$, the collision matrix $D$ is given by

$$D = \begin{bmatrix} C_0 D_0 & 0 & 0 \\ 0 & \ddots & 0 \\ 0 & 0 & C_{Q-1} D_{Q-1} \end{bmatrix}, \quad D_\alpha = \begin{bmatrix} h_\alpha^{eq}(\boldsymbol{x}_0,t) & 0 & 0 \\ 0 & \ddots & 0 \\ 0 & 0 & h_\alpha^{eq}(\boldsymbol{x}_{M-1},t) \end{bmatrix}, \quad (A.6)$$

where

$$h_\alpha^{eq}(\boldsymbol{x},t) = h_\alpha^{eq}(\boldsymbol{x},t)/T = w_\alpha \left(1 + \frac{\boldsymbol{e}_\alpha \cdot \boldsymbol{u}}{c_s^2} + \frac{(\boldsymbol{e}_\alpha \cdot \boldsymbol{u})^2}{2c_s^4} - \frac{\boldsymbol{u} \cdot \boldsymbol{u}}{2c_s^2} + B\delta t(\boldsymbol{e}_\alpha \cdot \nabla T)/T \right). \quad (A.7)$$

**Acknowledgements**

The research is partially supported by the National Natural Science Foundation of China (12202191) and Research Fund of State Key Laboratory of Mechanics and Control for Aerospace Structures (Nanjing University of Aeronautics and Astronautics) (MCAS-I-0324G04).

**Competing interests**

The authors declare no competing interests.